\documentclass[aps,pre,twocolumn,amsmath,amssymb,superscriptaddress,showpacs,longbibliography]{revtex4-1}

\usepackage{graphicx}
\usepackage{dcolumn}
\usepackage{bm}
\usepackage{mathrsfs}

\usepackage{natbib}
\usepackage[text={7in,9.5in},centering]{geometry}

\usepackage{hyperref}
\usepackage{url}

\usepackage{epsfig}
\usepackage{amsmath}
\usepackage{amssymb}
\usepackage{textcomp}

\usepackage{color}
\usepackage{float}

\usepackage{stackengine}
\usepackage{verbatim}

\usepackage{autonum}

\begin{document}

\title{Enhanced robustness of single-layer networks with redundant dependencies}

\author{G. Tim\'ar}
 \email[]{gtimar@ua.pt}
 \affiliation{Departamento de F\'\i sica da Universidade de Aveiro \& I3N, Campus Universit\'ario de Santiago, 3810-193 Aveiro, Portugal}

\author{Gy. Kov\'acs}
 \affiliation{Analytical Minds Ltd., 4933, Beregsur\'any, \'Arp\'ad Street 5., Hungary}

\author{J. F. F. Mendes}
 \affiliation{Departamento de F\'\i sica da Universidade de Aveiro \& I3N, Campus Universit\'ario de Santiago, 3810-193 Aveiro, Portugal}

\date{\today}

\begin{abstract}
Dependency links in single-layer networks offer a convenient way of modeling nonlocal percolation effects in networked systems where certain pairs of nodes are only able to function together. We study the percolation properties of the weak variant of this model: Nodes with dependency neighbours may continue to function if at least one of their dependency neighbours is active. We show that this relaxation of the dependency rule allows for more robust structures and a rich variety of critical phenomena, as percolation is not determined strictly by finite dependency clusters.
We study Erd\H os-R\'enyi and random scale-free networks with an underlying Erd\H os-R\'enyi network of dependency links.
We identify a special ``cusp'' point above which the system is always stable, irrespective of the density of dependency links.
We find continuous and discontinuous hybrid percolation transitions, separated by a tricritical point for Erd\H os-R\'enyi networks. For scale-free networks with a finite degree cutoff we observe the appearance of a critical point and corresponding double transitions in a certain range of the degree distribution exponent. We show that at a special point in the parameter space, where the critical point emerges, the giant viable cluster has the unusual critical singularity $S-S_c \propto (p-p_c)^{1/4}$. We study the robustness of networks where connectivity degrees and dependency degrees are correlated and find that scale-free networks are able to retain their high resilience for strong enough positive correlation, i.e., when hubs are protected by greater redundancy.
\end{abstract}

\maketitle

\section{Introduction}
\label{sec1}

The desire for an increasingly accurate description of networked systems has resulted in various useful generalizations of the classical percolation theory of random graphs. One branch of these generalizations stems from the notion that the functioning of a particular node may depend on the functioning of certain other nodes in the system to which the node in question may not be directly connected. This idea has led to the definition of mutually connected components in interdependent (or multiplex) networks \cite{buldyrev2010catastrophic, son2012percolation, baxter2012avalanche}. Such a network is composed of various network layers and a node in one layer may have various interdependency neighbours in other layers. According to the most common definition a mutually connected component is one that is connected on all layers, i.e., the interdependency neighbours of nodes in a connected cluster in one layer must also form connected clusters on all other layers. The giant mutually connected component shows an increased vulnerability to random damage, compared to the giant components of the individual layers and, for random uncorrelated layers, collapses in a discontinuous hybrid transition \cite{baxter2012avalanche}, such as the one also seen in, e.g., $k$-cores \cite{dorogovtsev2006k}.
Interdependent and multiplex networks have enjoyed considerable popularity in recent years due to their ability to model important and easily observed real-world systems such as online social networks, transportation networks, neuronal networks and many more \cite{boccaletti2014structure}.

The notion of dependencies between nodes in a network was exploited in a somewhat simpler generalization of ordinary percolation by Parshani et al. in Ref. \cite{parshani2011critical}. Here a single-layer network of connectivity links is considered, where there may also be ``dependency links'' between certain pairs of nodes. The percolation rule is given as a deactivation process. Initially a given fraction of nodes in the network is activated, and nodes can only remain active if (i) they belong to the giant connected component of active nodes and (ii) all of their dependency neighbours are also active. This deactivation process either leads to a stable situation, where a fraction of nodes in the network remain active, or all nodes in the network are deactivated. Using the fraction of initially active nodes as control parameter, it was shown in Ref. \cite{parshani2011critical} that the stable fraction of active nodes may undergo a continuous or a discontinuous transition, depending on the density and configuration of dependency links in the network. The continuous transition regime is essentially characterized by the fraction of nodes without dependencies and the discontinuous one by the distribution of the sizes of dependency groups (finite clusters of nodes connected by dependency links). Scale-free networks, that are highly robust in the continuous transitions regime, were found to be particularly fragile in the discontinuous transitions regime. The effect of various different dependency group size distributions is explored in Refs. \cite{bashan2011percolation, bashan2011combined, lin2017robustness}. Several generalizations of this model have been investigated in recent years, such as networks with directed \cite{niu2016percolation} and time-varying dependency links \cite{bai2016robustness} and multilayer networks with dependency links both between and within the individual layers \cite{liu2016cascading}.

The dependency rules assumed in most of the initial work on interdependent networks and single-layer networks with internal dependency links are too restrictive to describe certain systems that do not exhibit the predicted fragility. A relaxation of the standard multiplex percolation rule was explored in Refs. \cite{baxter2014weak, baxter2020exotic} where a node is defined to belong to a component if it has at least one neighbour on each layer in the same component---without the requirement that the component be connected on each layer. It was found that a two-layer network, of uncorrelated random networks, in this case still exhibits a continuous percolation transition, as opposed to the discontinuous hybrid transition of the standard giant mutually connected component. A different approach was studied in Ref. \cite{radicchi2017redundant} where mutually connected components in a multiplex network were required to be connected on at least two layers, as opposed to all of them. For numbers of layers greater than or equal to two, the addition of new layers in this case increases the robustness of the system.

In this paper we consider single-layer networks with internal dependency links and propose a relaxation of the percolation rule introduced in Ref. \cite{parshani2011critical}. We consider a node to belong to a component if it has at least one connectivity link to the given component and---if it has dependency links---at least one of its dependency neighbours is also in the same component. This weaker dependency rule may be suitable for modeling systems where individual nodes require some kind of input from other nodes to function, but this input may be supplied by various different nodes, not just one. This model has the interesting feature that the size of the giant component is non-monotonic as a function of the density of dependency links. Very few dependencies, as well as a high number of redundant dependencies both correspond to robust structures, with a ``valley'' of more fragile states in between. We investigate the percolation properties of such systems, with Erd\H os-R\'enyi and scale-free connectivity networks, where dependency links are placed randomly, i.e., the dependency network is Erd\H os-R\'enyi with a given mean dependency degree. We find continuous and discontinuous percolation transitions separated by a tricritical point for Erd\H os-R\'enyi connectivity networks. For scale-free networks with a finite degree cutoff we show that in a certain range of the degree distribution exponent a critical point appears, which is accompanied by a non-smooth switch between continuous and discontinuous transitions, as well as double percolation transitions. We show that at the point where a critical point appears the giant component has the critical singularity $S-S_c \propto (p-p_c)^{1/4}$. We also consider the situation where connectivity and dependency networks are correlated and find that robustness can be greatly improved by positive correlations between connectivity and dependency degrees.

The paper is organized as follows. In Section \ref{sec2} we introduce our model and discuss some implications and important differences compared to the definitions of Ref. \cite{parshani2011critical}. In Section \ref{sec3} we set up self-consistency equations to solve our model for uncorrelated random connectivity and dependency networks, and present numerical solutions compared with simulation results. In Section \ref{sec4} we explore the various possible forms of critical behaviour. We derive conditions for critical thresholds and obtain the order parameter exponent for the various cases analytically. In Section \ref{sec5} we present results for scale-free connectivity networks with a finite degree cutoff, explaining the origin of double percolation transitions and the unique order parameter exponent $\beta = 1/4$. In Section \ref{sec6} we study the effect of correlations between connectivity and dependency degrees. We give our conclusions in Section \ref{sec7}.

\section{Connected components in networks with redundant dependency links}
\label{sec2}

We consider an arbitrary undirected network of connectivity links between nodes and an arbitrary undirected network of dependency links between the same nodes. Connectivity links establish reachability relationships between nodes and serve as the backbone for connected components. Dependency links signify the conditions that certain nodes can only function if they are able to reach certain other nodes. Connected components in our percolation model of ``weak dependencies'' may be defined as the stable state of the following iterative process.

\begin{enumerate}
 \item Identify all connected components based on connectivity links.
 
 \item Remove all connectivity links of all nodes $i$ that have at least one dependency neighbour, and none of these dependency neighbours are in the connected component of node $i$.
 
 \item Repeat steps 1 and 2 until no further changes are made.
 
\end{enumerate}

\begin{figure}
\centering
\includegraphics[width=\columnwidth,angle=0.]{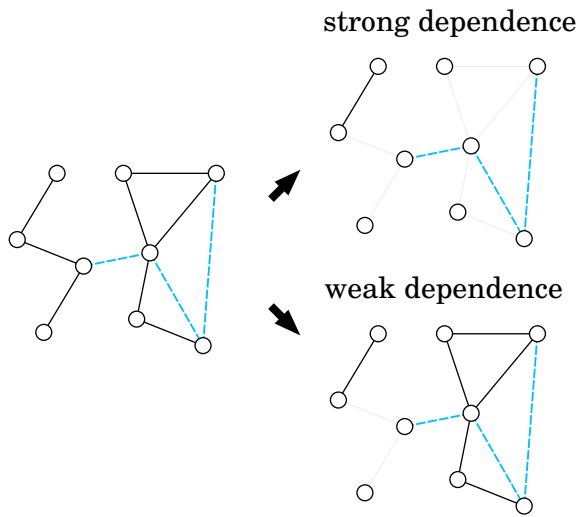}
\caption{Example of a simple network with dependencies consisting of two connected components (in the standard percolation sense). Solid black lines represent connectivity links, dashed blue lines represent dependency links. Strongly and weakly dependent components (of the same network) are shown on the right.}
\label{fig:schematic}
\end{figure}

\noindent
We will refer to the resulting components as \emph{weakly dependent components} (WDCs) to distinguish from ``ordinary'' connected components based purely on connectivity links. Similarly we will refer to the components defined in \cite{parshani2011critical} as \emph{strongly dependent components} (SDCs).
Note that in the strong dependence model Step 2 of the above iterative process is replaced with ``Remove all connectivity links of all nodes $i$ that have at least one dependency neighbour, and at least one of these dependency neighbours is not in the connected component of node $i$.'' A schematic representation of strongly and weakly dependent components is shown in Fig. \ref{fig:schematic} for a simple network of two connected components (in the standard percolation sense).

From the definitions it follows that the SDC that node $i$ belongs to is always a subgraph of its WDC. Also, the sizes of WDCs may increase or decrease as a result of the addition of a dependency link, while the sizes of SDCs cannot increase.
For large, sparse, random uncorrelated connectivity and dependency networks the giant connected component (GCC) of the dependency network constitutes a barrier to the existence of SDCs: none of the nodes that belong to the GCC of the dependency network can belong to an SDC. For this reason it is the finite clusters (dependency groups) of the dependency network that determine the percolation properties under the strong dependency rule \cite{parshani2011critical, bashan2011percolation, bashan2011combined, lin2017robustness}. Such a restriction does not apply in the case of the weak dependency rule. This might make the weak dependency model (or some combination of the weak and strong models) a better candidate to describe the behaviour of certain real-world systems with dependencies.
Similarly to $k$-cores and mutually connected components in multiplex networks, the probability that a random node belongs to a finite dependent component that contains dependencies is negligible in large, sparse, uncorrelated random networks, according to both the weak and strong definition. Finite components, with non-negligible probability, exist in both models only if none of the nodes in the given component have dependencies.

\begin{figure}
\centering
\includegraphics[width=\columnwidth,angle=0.]{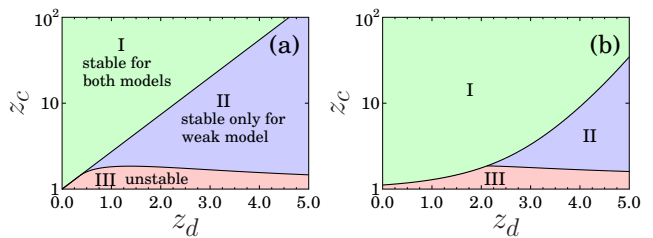}
\caption{Phase diagrams of (a) Erd\H os-R\'enyi and (b) scale-free connectivity networks with mean connectivity degree $z_c$, both with an Erd\H os-R\'enyi dependency network of mean degree $z_d$. The scale-free connectivity network is an uncorrelated random network with a degree distribution of the form $P_c(k) \sim (k + B)^{-\gamma}$, with lower and upper degree cutoffs $k_{\textrm{min}} = 1$ and $k_{\textrm{max}} = 1000$, respectively. The parameter $B$ was chosen to achieve a given mean degree $z_c$. The degree distribution exponent was $\gamma = 3$. The phase separation curves were obtained by numerical analysis of the self-consistency equations presented in Section \ref{sec3}.}
\label{fig:strong_weak}
\end{figure}

In this paper we focus on the properties of the giant weakly dependent component (GWDC) for Erd\H os-R\'enyi and random scale-free connectivity networks, with an Erd\H os-R\'enyi dependency network. To demonstrate the effect of the weak model (compared to the strong model), in Fig. \ref{fig:strong_weak} we present phase diagrams of these two types of connectivity networks, indicating the regions where a giant dependent component---strong or weak---exists. The threshold value of $z_c$ in the strong model is a monotonically increasing function of $z_d$, while it initially increases, then decreases in the weak model due to an increasing number of redundant dependency links. This qualitative behaviour applies to both types of connectivity networks. The weak model allows for stable structures in a much wider range of parameters. (Note the logarithmic scale on the $y$ axis.)

\section{Self-consistency equations}
\label{sec3}

For infinite sparse random uncorrelated connectivity and dependency networks, exploiting their local tree-likeness, the problem of finding the relative size of the GWDC may be solved by setting up appropriate self-consistency equations. Let us consider a connectivity network with degree distribution $P_c(k)$ and a dependency network with degree distribution $P_d(k)$. We consider each connectivity link to be active with probability $p$. We introduce two probabilities that will allow us to write exact self-consistency equations for this problem. First, let $x$ be the probability that following a random connectivity link in a random direction we can reach the GWDC. Second, let $y$ be the probability that we encounter a node in the GWDC by following a random dependency link emanating from a node in the GWDC.
We can write the following equation for $x$,

{
\medmuskip=0mu
\thinmuskip=0mu
\thickmuskip=0mu
\begin{align}
x = p \left[  1 - \sum_{k=1}^{\infty} \frac{k P_c(k)}{z_c} (1-x)^{k-1}  \right] \left[ 1 - \sum_{k=1}^{\infty} P_d(k) (1-y)^k  \right],
\label{eq3.10}
\end{align}
}

\noindent
where $z_c$ denotes the mean degree of the connectivity network. The first factor in square brackets in Eq. (\ref{eq3.10}) gives the probability that following a random connectivity link, the node encountered has at least one outgoing connectivity link to the GWDC. The second factor gives the probability that following a random connectivity link, the node encountered has either no dependency neighbours, or has at least one dependency neighbour in the GWDC. These two factors must be multiplied by $p$, the probability that the link on which we arrived is active. The equation for $y$ is simpler,

\begin{align}
y = \left[ 1 - \sum_{k=0}^{\infty} P_c(k) (1-x)^k  \right].
\label{eq3.20}
\end{align}

\noindent
Note that the right-hand side of Eq. (\ref{eq3.20}) does not depend on $y$, i.e., $x$ is the only independent variable in this problem.
We can solve Eqs. (\ref{eq3.10}) and (\ref{eq3.20}) numerically by iteration. Once the solutions $x$ and $y$ are found, we can express the relative size of the GWDC,

{
\medmuskip=0mu
\thinmuskip=0mu
\thickmuskip=0mu
\begin{align}
S = \left[  1 - \sum_{k=0}^{\infty} P_c(k) (1-x)^k  \right] \left[ 1 - \sum_{k=1}^{\infty} P_d(k) (1-y)^k  \right].
\label{eq3.30}
\end{align}
}

The self-consistency equations (\ref{eq3.10}) and (\ref{eq3.20}) can be written in more compact form using probability generating functions.
Let us introduce the generating function for the connectivity degree distribution, the dependency degree distribution, and the connectivity ``excess degree'' distribution, respectively,

\begin{align}
\label{eq3.40}
G_c(x) &= \sum_{k=0}^{\infty} P_c(k) x^k,\\
\label{eq3.50}
G_d(x) &= \sum_{k=0}^{\infty} P_d(k) x^k,\\
\label{eq3.60}
H_c(x) &= \sum_{k=0}^{\infty} \frac{(k+1) P_c(k+1)}{z_c} x^k.
\end{align}

\noindent
Using these generating functions, Eqs. (\ref{eq3.10}) and (\ref{eq3.20}) can be written as

\begin{align}
\label{eq3.70}
x &= p \left[ 1 - H_c(1-x)  \right]   \left[  1 - G_d(1-y) + P_d(0)   \right],\\
\label{eq3.80}
y &= \left[ 1 - G_c(1-x)  \right].
\end{align}

\noindent
Substituting Eq. (\ref{eq3.80}) into Eq. (\ref{eq3.70}) we arrive at one single self-consistency equation for $x$,

{
\medmuskip=0mu
\thinmuskip=0mu
\thickmuskip=0mu
\begin{align}
x \, &= \,  p \left[ 1 - H_c(1-x)  \right]   \left[  1 + P_d(0) - G_d(G_c(1-x) )   \right] \nonumber \\
\label{eq3.95}
&\equiv \, p\Psi(x).
\end{align}
}

\noindent
The relative size of the GWDC may be expressed as

{
\medmuskip=0mu
\thinmuskip=0mu
\thickmuskip=0mu
\begin{align}
S = \left[ 1 - G_c(1-x)  \right]   \left[  1 + P_d(0) - G_d(G_c(1-x) )   \right].
\label{eq3.100}
\end{align}
}

\begin{figure}
\centering
\includegraphics[width=\columnwidth,angle=0.]{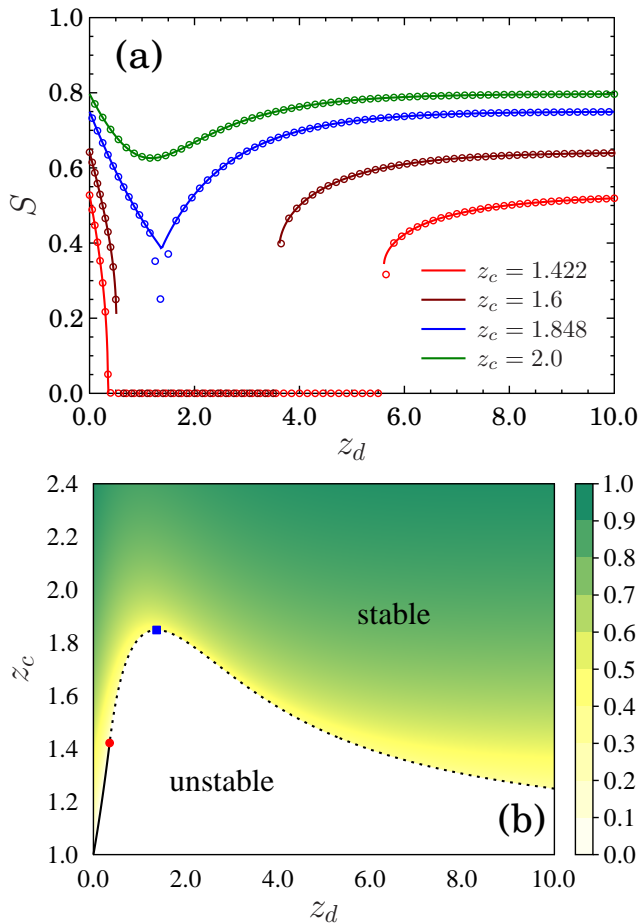}
\caption{(a) Relative size $S$ of the GWDC as a function of mean dependency degree $z_d$, where both the connectivity and dependency networks are Erd\H os-R\'enyi networks. Curves are shown for four different values of the mean connectivity degree $z_c$. Open circles represent simulation results, solid lines correspond to the numerical solution of Eqs. (\ref{eq3.95}) and (\ref{eq3.100}). (Number of nodes was $N=10^6$ in all cases and results were averaged over $100$ realizations.) (b) Phase diagram of the same network class, with $S$ overlaid as a colormap. Solid and dashed black lines correspond to continuous and discontinuous transitions respectively. The tricritical point and cusp point (see Sections \ref{sec41} and \ref{sec42}) are marked by a solid red circle and solid blue square, respectively.}
\label{fig:intro_fig}
\end{figure}

Figure \ref{fig:intro_fig}(a) shows numerical solutions for $S$ obtained using Eqs. (\ref{eq3.95}) and (\ref{eq3.100}), compared with simulations. Both the connectivity and dependency networks are Erd\H os-R\'enyi, with mean connectivity degree $z_c$ and mean dependency degree $z_d$. For high enough $z_c$ [$z_c = 2$ in Fig. \ref{fig:intro_fig}(a)] we see that the GWDC exists in the entire range of $z_d$ values, although its size exhibits a minimum for an intermediate value of $z_d$. With decreasing $z_c$, below the point $z_c \approx 1.848$, the $S(z_d)$ curve breaks up into two separate regions that correspond to a low and high density of dependencies.  The region in the middle is not able to support a GWDC. It can also be seen that, with increasing $z_d$, the GWDC disappears and reappears in a discontinuous transition [$z_c=1.6$ in Fig. \ref{fig:intro_fig}(a)]. At the point $z_c \approx 1.422$ the first transition changes to continuous. All four numerical curves show very good agreement with simulations except close to the critical regions, where large fluctuations are expected. To further demonstrate the different types of transitions, Figure \ref{fig:intro_fig}(b) shows a phase diagram of the same network class, with $S$ overlaid as a colormap.

For the sake of completeness we present here also the exact self-consistency equations necessary to solve the strong variant of the model, in the same settings. Let the probabilities $x$ and $y$ have the same meaning as above (except for replacing the word ``GWDC'' with ``GSDC'' in the definition). The two equations for the strong model are

\begin{align}
x = p \left[  1 - \sum_{k=1}^{\infty} \frac{k P_c(k)}{z_c} (1-x)^{k-1}  \right] \left[ \sum_{k=0}^{\infty} P_d(k) y^k  \right],
\label{eq3.110}
\end{align}

\begin{align}
y = \left[ 1 - \sum_{k=0}^{\infty} P_c(k) (1-x)^k \right] \left[  \sum_{k=1}^{\infty} \frac{k P_d(k)}{z_d} y^{k-1}  \right],
\label{eq3.120}
\end{align}

\noindent
and the relative size of the GSDC is expressed as

\begin{align}
S = \left[  1 - \sum_{k=0}^{\infty} P_c(k) (1-x)^k  \right] \left[ \sum_{k=0}^{\infty} P_d(k) y^k  \right].
\label{eq3.130}
\end{align}

\section{Critical behaviour}
\label{sec4}

We explore the various types of critical behaviour that occur in the weak dependency model, associated with the appearance of the GWDC.
To study the behaviour of Eq. (\ref{eq3.95}) it will be useful to introduce the function $f(x) = \Psi(x) / x$. (We should remember that apart from $x$, $f$ depends on the distributions $P_c$ and $P_d$.)
Equation (\ref{eq3.95}) now reads

\begin{align}
pf(x) = 1.
\label{eq4.10}
\end{align}

\noindent
Note that Eq. (\ref{eq4.10}) is equivalent to Eq. (\ref{eq3.95}) only for $x>0$. $x = 0$ is always a solution of Eq. (\ref{eq3.95}). Apart from this trivial solution, all other solutions can be found using Eq. (\ref{eq4.10}). The first nonzero solution $x^*$ occurs when the maximum of curve $pf(x)$ first becomes $1$. If the maximum of $f$ in the range $]0, 1]$ is denoted by $f_{\textrm{max}}$, then the value of $p$ at which this happens is given by $p_c = 1/f_{\textrm{max}}$. We now discuss the various possible situations that correspond to different types of phase transitions with different critical singularities.

\subsection{Continuous transitions, discontinuous transitions and tricritical point}
\label{sec41}

\begin{figure}[H]
\centering
\includegraphics[width=\columnwidth,angle=0.]{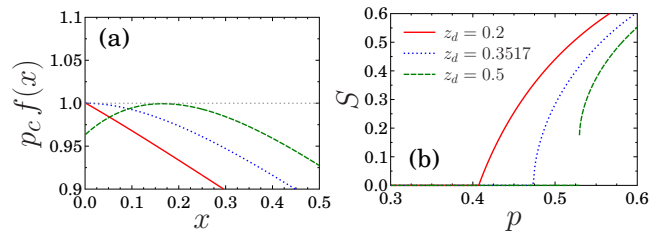}
\caption{(a) Function $f(x)$ (scaled to be at the critical threshold) for different values of the mean dependency degree $z_d$. (b) Relative size $S$ of the GWDC as a function of connectivity link activation probability $p$, for different values of $z_d$. The three curves on each panel correspond to a continuous transition ($z_d = 0.2$), the tricritical point ($z_d = 0.3517$) and a discontinuous transition ($z_d = 0.5$). The mean connectivity degree is $z_c = 3$ in all cases. Both the connectivity and the dependency network is Erd\H os-R\'enyi.}
\label{fig:cont_tri_disc}
\end{figure}

Let us first consider the situation where $f(x)$ is monotonically decreasing, i.e., $f'(x) < 0$ for all $x > 0$ [solid red line in Fig. \ref{fig:cont_tri_disc}(a)]. In this case $f_{\textrm{max}} = \lim_{x \to 0} f(x)$, which corresponds to a continuous transition. [Remember that $f(0)$ is not defined.]

\noindent
This limit can be evaluated using L'Hospital's rule,

\begin{align}
\lim_{x \to 0} f(x) = \lim_{x \to 0} \frac{\Psi(x)}{x} = \lim_{x \to 0} \frac{\Psi'(x)}{x'} = \Psi'(0).
\label{eq41.10}
\end{align}

\noindent
Using Eqs. (\ref{eq3.40}), (\ref{eq3.50}), (\ref{eq3.60}), (\ref{eq3.95}) and basic properties of probability generating functions we obtain

\begin{align}
\Psi'(0) = P_d(0) \frac{\langle k(k-1) \rangle_c}{\langle k \rangle_c},
\label{eq41.20}
\end{align}

\noindent
resulting in the threshold for a continuous phase transition,

\begin{align}
p_c = \frac{ \langle k \rangle_c }{ P_d(0) \langle k(k-1) \rangle_c }.
\label{eq41.30}
\end{align}

\noindent
We see that only $P_d(0)$ plays a role, i.e., the shape of the dependency degree distribution is irrelevant and only the fraction of nodes with no dependencies matters.
For the case where both the connectivity and dependency networks are Erd\H os-R\'enyi (with mean degrees $z_c$ and $z_d$, respectively) and $p_c = 1$, the condition (\ref{eq41.30}) can be written simply as

\begin{align}
z_c = e^{z_d}.
\label{eq41.40}
\end{align}

\noindent
The corresponding curve is plotted in Fig. \ref{fig:intro_fig}(b) as a solid black line.
Using the probability $p$ as a control parameter, the behaviour of $S$ near the transition is given by $S \propto (p-p_c)^{1}$ [see solid red line in Fig. \ref{fig:cont_tri_disc}(b)]. This can be shown by expanding Eq. (\ref{eq3.95}) about the point $(x=0, p=p_c)$ and using Eq. (\ref{eq4.10}) (see Appendix for details).

As noted above, a continuous transition can only happen if $f_{\textrm{max}} = \lim_{x \to 0} f(x)$. If $\lim_{x \to 0} f'(x) > 0$ then this is not the case and $f_{\textrm{max}} = f(x^*)$ for some $x^*>0$, meaning that the nontrivial solution emerges with a jump [see dashed green line in Fig. \ref{fig:cont_tri_disc}(a,b)]. The condition for this to happen is $f'(x^*) = 0$, and the corresponding threshold for the discontinuous transition is

\begin{align}
p_c = \frac{1}{f(x^*)}.
\label{eq41.50}
\end{align}

\noindent
The behaviour of $S$ near such a transition is given by $S-S_c \propto (p-p_c)^{1/2}$. This type of phase transition is also frequently referred to as a hybrid transition, involving a discontinuity and a critical singularity. The critical behaviour can be derived by expanding Eq. (\ref{eq3.95}) about the point $(x=x^*, p=p_c)$ and using the condition $f'(x^*) = 0$ (see Appendix for details).

Assuming for now that $f$ has only one maximum in the interval $]0, 1]$, we saw that $\lim_{x \to 0} f'(x) < 0$ results in a continuous transition, while $\lim_{x \to 0} f'(x) > 0$ produces a discontinuous hybrid transition. The two types of transitions meet at a tricritical point where $\lim_{x \to 0} f'(x) = 0$ [see dotted blue line in Fig. \ref{fig:cont_tri_disc}(a,b)]. Using L'Hospital's rule we find

{
\medmuskip=0mu
\thinmuskip=0mu
\thickmuskip=0mu
\begin{align}
\lim_{x \to 0} f'(x) &= \lim_{x \to 0} \left( \frac{\Psi(x)}{x} \right)' = \lim_{x \to 0} \frac{x \Psi'(x) - \Psi(x)}{x^2} = \nonumber \\
&= \lim_{x \to 0} \frac{(x \Psi'(x) - \Psi(x))'}{(x^2)'} = \frac{\Psi''(0)}{2}.
\label{eq41.60}
\end{align}
}

\noindent
Using the properties of generating functions, $\Psi''(0)$ can be expressed as

\begin{align}
\Psi''(0) = 2 \langle k(k-1) \rangle_c  \langle k \rangle_d - P_d(0) \frac{ \langle k(k-1)(k-2) \rangle_c }{ \langle k \rangle_c },
\label{eq41.70}
\end{align}

\noindent
resulting in the condition for a tricritical point,

\begin{align}
\frac{P_d(0)}{2 \langle k \rangle_d} = \frac{ \langle k(k-1) \rangle_c  \langle k \rangle_c }{ \langle k(k-1)(k-2) \rangle_c }.
\label{eq41.80}
\end{align}

\noindent
For Erd\H os-R\'enyi networks, and setting $p_c = 1$, Eqs. (\ref{eq41.80}) and (\ref{eq41.30}) are equivalent to

\begin{align}
z_c &= e^{1 / (2 z_c)} \nonumber \\
z_d &= 1 / (2 e^{z_d}),
\label{eq41.90}
\end{align}

\noindent
which have the solution $z_c^{\textrm{tc}} \approx 1.4215$ and $z_d^{\textrm{tc}} \approx 0.3517$. This tricritical point is shown as a solid red circle on Fig. \ref{fig:intro_fig}(b).
Using the probability $p$ as a control parameter, the behaviour of $S$ near such a transition is given by $S \propto (p-p_c)^{1/2}$. Thus the tricritical point has the same type of critical singularity as hybrid transitions, only here the jump size is zero. The critical behaviour can be derived by expanding Eq. (\ref{eq3.95}) about the point $(x=0, p=p_c)$ and using the condition $\lim_{x \to 0} f'(x) = 0$ (see Appendix for details). Equation (\ref{eq41.80}) also means that infinite scale-free connectivity networks with a degree distribution exponent $\gamma < 4$ (and no finite degree cutoff) cannot have a tricritical point for any finite value of the mean dependency degree [as long as $P_d(0) > 0$].

\subsection{``Cusp'' point}
\label{sec42}

\begin{figure}[H]
\centering
\includegraphics[width=\columnwidth,angle=0.]{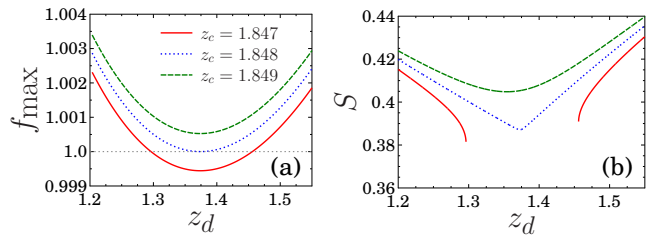}
\caption{(a) The maximum (with respect to $x$) of the function $f(x, z_d)$, as a function of $z_d$, for three different values of $z_c$. (b) The relative size $S$ of the GWDC as a function of $z_d$ for the same three values of $z_c$ as in panel (a). The three curves in each panel correspond to a case with two discontinuous transitions with $S=0$ in between ($z_c = 1.847$), the cusp point ($z_c = 1.848$) and a situation with no transitions ($z_c = 1.849$). Both the connectivity and the dependency network are Erd\H os-R\'enyi.}
\label{fig:cusp}
\end{figure}

Assuming an Erd\H os-R\'enyi dependency network, let us now set $p = 1$ and take $z_d$, the mean dependency degree, to be our control parameter. As can be seen in Fig. \ref{fig:intro_fig}, for Erd\H os-R\'enyi connectivity networks, for small enough values of $z_c$ there are two percolation transitions, when varying $z_d$. The GWDC first disappears, and then reappears for large enough $z_d$. When $z_c<z_c^{\textrm{tc}}$, the first transition of the two is continuous and for $z_c>z_c^{\textrm{tc}}$ it is discontinuous ($z_c^{\textrm{tc}}$ denoting the value of $z_c$ at the tricritical point).
An interesting consequence of the weak dependency model is that for large enough $z_c$, percolation transitions disappear altogether, i.e., a GWDC always exists, for any $z_d$. For the particular value of $z_c$ where this happens, the $S(z_d)$ curve has a unique ``cusp'' shape.
To find the values of $z_c$ and $z_d$ where this cusp point occurs, let us consider also the $z_d$-dependence of the function $f$: $f = f(x,z_d)$. It is easy to see that the cusp point is a saddle point of the function $f(x,z_d)$: at this point $f$ is maximal with respect to $x$ and minimal with respect to $z_d$ [see Fig. \ref{fig:cusp}(a)]. The cusp point, therefore, has the following conditions,

\begin{align}
&\frac{\partial f}{\partial x}  \biggr|_{x^{\textrm{cp}}, z_d^{\textrm{cp}}} = \frac{\partial f}{\partial z_d}  \biggr|_{x^{\textrm{cp}}, z_d^{\textrm{cp}}} = 0. \nonumber \\
&f(x^{\textrm{cp}}, z_d^{\textrm{cp}}) = 1.
\label{eq42.10}
\end{align}

\noindent
These are three equations for the three unknowns $x^{\textrm{cp}}$, $z_d^{\textrm{cp}}$ and $z_c^{\textrm{cp}}$. For Erd\H os-R\'enyi networks the cusp point occurs at $z_c^{\textrm{cp}} \approx 1.848$ and $z_d^{\textrm{cp}} \approx 1.371$. Using $z_d$ as control parameter the behaviour of $S$ near the cusp point is given by $S-S^{\textrm{cp}} \propto |z_d-z_d^{\textrm{cp}}|^1$ (for $z_c = z_c^{\textrm{cp}}$), i.e., the critical exponent changes from $1/2$ to $1$ at this point. Note also that $S-S^{\textrm{cp}}$ is proportional to the distance from $z_d^{\textrm{cp}}$ on both sides of the cusp point [see Fig. \ref{fig:cusp}(b)]. This critical behaviour can be derived by expanding Eq. (\ref{eq3.95}) about the point $(x=x^{\textrm{cp}}, z_d=z_d^{\textrm{cp}})$ and using the conditions (\ref{eq42.10}) (see Appendix for details).

\subsection{Critical point and double transitions}
\label{sec43}

\begin{figure}[H]
\centering
\includegraphics[width=\columnwidth,angle=0.]{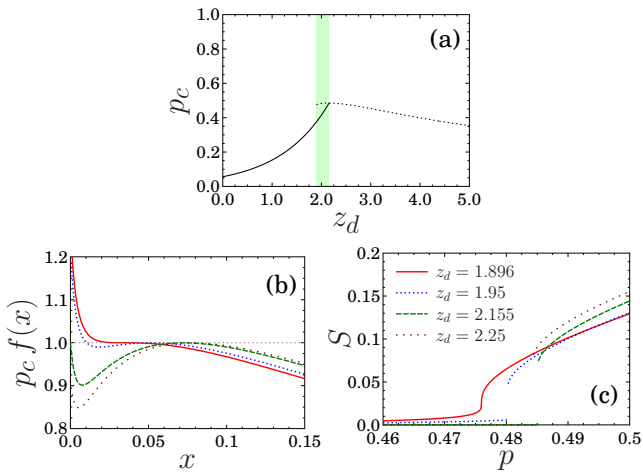}
\caption{(a) Phase diagram for a scale-free connectivity network ($z_c = 3$, $\gamma = 3$, $k_{\textrm{min}} = 1$, $k_{\textrm{max}} = 1000$) and Erd\H os-R\'enyi dependency network. Solid line represents continuous transitions, dashed line corresponds to discontinuous transitions. The shaded green region indicates double percolation transitions. (b) Function $f(x)$ (scaled to be at the critical threshold) for different values of the mean dependency degree $z_d$. (c) Relative size $S$ of the GWDC as a function of connectivity link activation probability $p$, for different values of $z_d$. The four curves in panels (b) and (c) correspond to the critical point ($z_d = 1.896$), a double transition ($z_d = 1.95$), the point where continuous transitions disappear ($z_d = 2.155$) and a discontinuous transition ($z_d = 2.25$).}
\label{fig:double_trans}
\end{figure}

Up to now we have assumed that the function $f$ has one maximum in the interval $]0, 1]$. This is true for Erd\H os-R\'enyi connectivity networks, but not necessarily so for broader degree distributions. We consider scale-free connectivity networks with degree distributions of the form

\begin{align}
P_c(k) = A (k + B)^{-\gamma},
\label{eq43.10}
\end{align}

\noindent
with finite lower and upper degree cutoffs $k_{\textrm{min}}$ and $k_{\textrm{max}}$. The parameter $A$ is a normalization constant and $B$ is adjusted in order to achieve a given mean connectivity degree $z_c$. Figure \ref{fig:double_trans}(a) shows the phase diagram of a network with a connectivity degree distribution of the form (\ref{eq43.10}) with $z_c = 3$, $\gamma = 3$, $k_{\textrm{min}} = 1$ and $k_{\textrm{max}} = 1000$. The diagram was obtained by numerical solution of Eq. (\ref{eq3.95}). (The dependency degree distribution is Erd\H os-R\'enyi with mean dependency degree $z_d$, as before.) The transition is continuous for low values and discontinuous for high values of $z_d$, similar to the case of Erd\H os-R\'enyi connectivity networks. However, here there is no smooth switch between the two types of transitions and the discontinuous line ends in a critical point. Also, in this case there is a region (shaded green on Fig. \ref{fig:double_trans}(a)) where double percolation transitions occur.

\noindent
For low values of $z_d$ the function $f$ has only one maximum (at $x \to 0$), corresponding to a continuous transition as discussed already in Section \ref{sec41}. With increasing $z_d$ the function $f$ develops a second local maximum at some point $x^* > 0$ [see solid red line in Fig. \ref{fig:double_trans}(b)], corresponding to a singularity on the $S(p)$ curve [solid red line in Fig. \ref{fig:double_trans}(c)]. The conditions for this point (a critical point) are

\begin{align}
f'(x^*) = f''(x^*) = 0.
\label{eq43.20}
\end{align}

\noindent
Expanding Eq. (\ref{eq3.95}) about the point $(x = x^*, p = p_c)$ [with $p_c = 1 / f(x^*)$] and using the conditions (\ref{eq43.20}) we find the behaviour $S - S_c \propto (p-p_c)^{1/3}$ close to the critical point (see Appendix for details). This type of singularity has been shown to appear in heterogeneous threshold models, see e.g. \cite{baxter2011heterogeneous, min2018competing}.

Increasing $z_d$ beyond the critical point we find that $f$ has two maxima, one at $x \to 0$ and one at some $x^* > 0$. The first maximum corresponds to a continuous transition, and the second corresponds to a subsequent discontinuous transition [see dotted blue line in Fig. \ref{fig:double_trans}(b,c)]. The second, discontinuous transition has the critical singularity $S-S_c \propto (p-p_c)^{1/2}$ already described in Section \ref{sec41}. The double transition exists in the region where $f$ has two local maxima and $\lim_{x \to 0} f(x) > f(x^*)$, where $x^*$ is the position of the second maximum. This region is shown shaded green in Fig. \ref{fig:double_trans}(a). The thresholds of the two transitions become equal when $\lim_{x \to 0} f(x) = f(x^*)$ [dashed green line in Fig. \ref{fig:double_trans}(b,c)], and above this point only discontinuous transitions can happen [dotted maroon line in Fig. \ref{fig:double_trans}(b,c)], as the local maximum at $x \to 0$ now corresponds to a non-physical solution.

Similar double transitions were also found in a mixed contagion model of simple and complex contagion \cite{min2018competing}. The situation is similar in our case: the continuous transition signifies an ordinary percolation phase, determined by the fraction of nodes without dependencies. The second, discontinuous transition corresponds to a ``complex'' phase, where at a certain point enough nodes have enough redundant dependencies to participate in a larger giant weakly dependent component.

\subsection{Emergence of the critical point}
\label{sec44}

For scale-free connectivity networks---i.e. ones with a degree distribution of the form (\ref{eq43.10})---we find that a critical point, and corresponding double transitions, only exist in a certain range $[\gamma^{\textrm{(low)}}, \gamma^{\textrm{(high)}}]$ of the degree distribution exponent.
For $z_c = 3$, $k_{\textrm{min}} = 1$ and $k_{\textrm{max}}=1000$ we find numerically that $\gamma^{\textrm{(low)}} \approx 2.146$ and $\gamma^{\textrm{(high)}} \approx 6.33$ (to a precision of $0.001$).
Starting at high values of $\gamma$ (coming from narrower connectivity degree distributions) the first point at which a critical point appears is $(\gamma = \gamma^{\textrm{(high)}}, z_d = z_d^{\textrm{(high)}}, p_c = p_c^{\textrm{(high)}} )$ for some $z_d^{\textrm{(high)}}$ and $p_c^{\textrm{(high)}}$. It can be shown that at this point $S$ has the critical behaviour $S \propto (p-p_c^{\textrm{(high)}})^{1/3}$ (see Appendix for derivation). At the other extreme, however, at the point $(\gamma = \gamma^{\textrm{(low)}}, z_d = z_d^{\textrm{(low)}}, p_c = p_c^{\textrm{(low)}} )$ we find

\begin{align}
S - S_c^{\textrm{(low)}} \propto (p-p_c^{\textrm{(low)}})^{1/4}.
\label{eq44.10}
\end{align}

\noindent
This unusual critical behaviour is explained in Section \ref{sec5}.

\section{Results for scale-free networks}
\label{sec5}

\begin{figure}[H]
\centering
\includegraphics[width=\columnwidth,angle=0.]{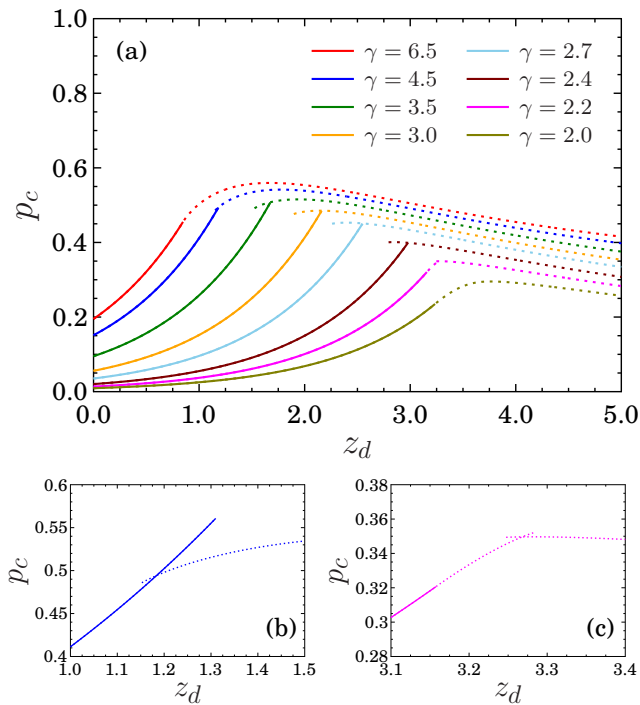}
\caption{(a) Phase diagrams for scale-free connectivity networks ($z_c = 3$, $k_{\textrm{min}} = 1$, $k_{\textrm{max}} = 1000$) for different values of the degree distribution exponent $\gamma$. Continuous transitions are represented by solid lines and discontinuous ones by dashed lines. A critical point and double transitions appear for intermediate values of $\gamma$, while there is a smooth switch (at a tricritical point) between continuous and discontinuous transitions for low and high $\gamma$ values. (b,c) Parts of two phase diagrams zoomed in from panel (a): $\gamma = 4.5$ (b) and $\gamma = 2.2$ (c). In panels (b,c) non-physical solutions of Eq. (\ref{eq3.95}) and (\ref{eq3.100}) are also shown.}
\label{fig:SF_gammas_together}
\end{figure}

As we saw in Section \ref{sec41}, Erd\H os-R\'enyi connectivity networks exhibit a tricritical point where continuous transitions switch smoothly to discontinuous ones. For scale-free connectivity networks this switch may be non-smooth for certain values of $\gamma$, due to the existence of a critical point. A critical point and corresponding double percolation transitions occur because the function $f(x)$ has two maxima (with respect to $x$) in a certain range of $z_d$ and $\gamma$ values.
Figure \ref{fig:SF_gammas_together}(a) shows phase diagrams for scale-free connectivity networks with different values of $\gamma$. For relatively small and relatively large $\gamma$ values the switch between continuous and discontinuous transitions is smooth, as for Erd\H os-R\'enyi connectivity networks. A critical point exists only for intermediate $\gamma$ values.

It is important to note that the networks considered here have a finite degree cutoff, therefore all moments of the degree distribution are finite. We are, therefore, not discussing the effect of asymptotically power-law degree distributions. We are using bounded power-law degree distributions to study the effect of broad distributions, that may also occur in real-world networks, where, of course, a finite cutoff always exists.

To understand how the critical point emerges it is useful to first look at zoomed-in versions of the phase diagrams for $\gamma = 4.5$ [Fig. \ref{fig:SF_gammas_together}(b)] and $\gamma = 2.2$ [Fig. \ref{fig:SF_gammas_together}(c)]. (In these figures the transition lines corresponding to the initial maximum are continued as long as this initial maximum exists, however they do not represent physical solutions after the intersection with the second transition line.)
Figure \ref{fig:SF_gammas_together}(c) shows that for $\gamma = 2.2$ a tricritical point actually exists and the critical point occurs in the discontinuous range, corresponding to discontinuous-discontinuous double transitions.
Considering that the jump size in discontinuous transitions goes to zero at the critical point, we know that close to the point where the critical point emerges the two local maxima---of $f(x)$---must be close together. In other words, the second local maximum, and the corresponding critical point, is a result of the initial maximum splitting into two local maxima.
Figures \ref{fig:SF_gammas_together} (b,c) suggest that the critical point emerges differently for high and low $\gamma$.

\begin{figure}[H]
\centering
\includegraphics[width=\columnwidth,angle=0.]{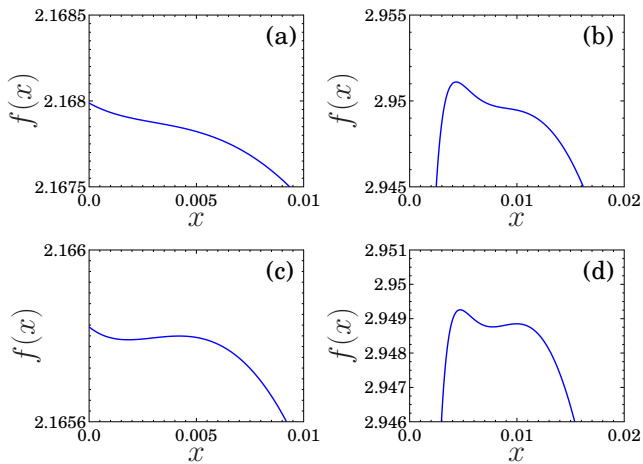}
\caption{Appearance of the second maximum of the function $f(x)$ for scale-free networks ($z_c = 3$, $k_{\textrm{min}} = 1$, $k_{\textrm{max}} = 1000$). Panels (a,c) show the appearance of the second maximum at $x=0$ [$\gamma = 6.1$, $z_d = 0.888$ and $z_d = 0.889$ for (a) and (c), respectively]. Panels (b,d) show the appearance of the second maximum at some $x>0$ [$\gamma = 2.15$, $z_d = 3.357$ and $z_d = 3.359$ for (b) and (d), respectively].}
\label{fig:max_birth}
\end{figure}

According to numerical analysis of the function $f(x)$, the highest $\gamma$ value for which $f(x)$ has two maxima for some $z_d$, i.e. a critical point exists, is $\gamma^{\textrm{(high)}} \approx 6.33$. The lowest such $\gamma$ value is $\gamma^{\textrm{(low)}} \approx 2.146$. (These values apply to degree cutoffs $k_{\textrm{min}} = 1$, $k_{\textrm{max}} = 1000$.)
We find that the emergence of the second maximum at $\gamma^{\textrm{(high)}}$ corresponds to the tricritical point ``breaking up'', i.e., happens at $x^*=0$. At $\gamma^{\textrm{(low)}}$ the second maximum is a result of the initial maximum (at some $x^* > 0$) splitting into two. These two different ways in which the second maximum can emerge are explained graphically in Fig. \ref{fig:max_birth}.

\noindent
The fact that at $\gamma^{\textrm{(low)}}$ the second maximum emerges at some $x^* > 0$ means that the following conditions must all hold,

\begin{align}
f'(x^*) = f''(x^*) = f'''(x^*) = 0,
\label{eq5.10}
\end{align}

\noindent
unlike at $\gamma^{\textrm{(high)}}$, where the emergence of the second maximum only has the conditions,

\begin{align}
f'(0) = f''(0) = 0.
\label{eq5.20}
\end{align}

\noindent
Having a zero-condition also for the third derivative of $f$ results in an unusual critical behaviour at $\gamma^{\textrm{(low)}}$. Expanding Eq. (\ref{eq3.95}) about the point $(x = x^*, p = p_c)$ and using the conditions (\ref{eq5.10}) leads to

\begin{align}
S - S_c \propto (p-p_c)^{1/4},
\label{eq5.30}
\end{align}

\noindent
(see Appendix for details).
This type of transition, in our current setup, has little practical significance, however, as the emergence of the second maximum at $\gamma^{\textrm{(low)}}$ happens at an $x^*$, which, although clearly positive, is still quite close to zero. Also, the maximum of $f$ is very sharp, and $f'''(x^*) \approx 0$ holds only in a very close vicinity of $x^*$, making the type of singularity difficult to observe. It is nevertheless an interesting phenomenon and its analysis sheds light on the necessary ingredients for this unique critical behaviour to occur.
Here we considered a scale-free connectivity degree distribution with a finite upper cutoff, therefore the double transitions and the unique critical exponent are not a consequence of power-law asymptotics, and would probably also occur for different types of broad degree distributions in this weak dependency model. Investigating the effect of ``truly'' scale-free networks, with no degree cutoff in the infinite network size limit, is left for future work.

\begin{figure}[H]
\centering
\includegraphics[width=\columnwidth,angle=0.]{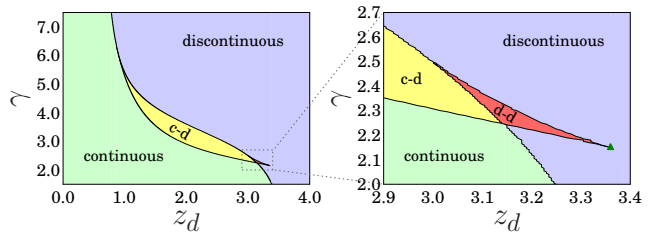}
\caption{Map of the different types of phase transitions occurring in a model of random scale-free connectivity network coupled with an Erd\H os-R\'enyi dependency network. The scale-free connectivity network had a degree distribution of the form of Eq. (\ref{eq43.10}), the parameter $B$ always adjusted to achieve a mean connectivity degree of $z_c = 3$. The green (blue) region corresponds to a single continuous (discontinuous) transition. The yellow (red) region corresponds to a double transition of the type ``continuous-discontinuous'' (``discontinuous-discontinuous''). The point marked with a green triangle corresponds to the conditions (\ref{eq5.10}) and the critical behaviour (\ref{eq5.30}). Right panel is a zoomed-in version of the framed rectangular area of the left panel.}
\label{fig:pt_map}
\end{figure}

Figure \ref{fig:pt_map} presents a map of the complex landscape of various types of transitions that are possible in the scale-free networks considered here. The boundary curves were determined numerically.

\section{Effect of correlation between connectivity and dependency degree}
\label{sec6}

Degree-degree correlations are a common feature among many real-world networks, both natural and artificial. In single-layer networks, without dependencies, positive (assortative) correlations between nearest neighbour degrees increase the robustness of networks \cite{goltsev2008percolation}, i.e., decrease the percolation threshold in random percolation (in terms of the fraction of active links). Similar results were found for multiplex networks \cite{min2014network, reis2014avoiding, bianconi2018multilayer}: positive correlations between degrees of nodes on different layers made networks more robust against random damage.
Here we investigate the effect of positive correlations between the connectivity and dependency degrees of nodes on the weakly dependent percolation threshold. There are various ways in which such a correlated network model could be built. As before we want to start with an uncorrelated random connectivity network with arbitrary degree distribution $P_c(k)$. We also fix the mean dependency degree, $z_d$, which may be different from $z_c$. To allow for simple, exact, self-consistency equations (in the infinite network size limit) we assume that the distribution of dependency degrees for a node of given connectivity degree is given by

{
\medmuskip=0mu
\thinmuskip=0mu
\thickmuskip=0mu
\begin{align}
P(k_d = k' | k_c = k) = (1-f) \frac{z_d^{k'} e^{-z_d}}{k'!} + f \frac{ (k\frac{z_d}{z_c})^{k'} e^{- (k\frac{z_d}{z_c}) }}{k'!},
\label{eq6.10}
\end{align}
}

\noindent
where $f$ is a tuning parameter: it allows us to interpolate between an Erd\H os-R\'enyi dependency network which has no correlation with the connectivity network ($f = 0$), and a network in which a node's dependency degree is Poisson distributed, with the node's rescaled connectivity degree as the mean (maximal correlation, $f = 1$). This kind of network construction resembles the ``hidden variable'' model of Chung and Lu \cite{chung2002connected, chung2002average}, where a unique degree distribution is prescribed for each node, parametrized by the expected degree, a hidden variable of the given node. It is easy to check that the mean degree of the dependency network, using Eq. (\ref{eq6.10}), is indeed equal to $z_d$, for any $f$.
(Note that $f = 1$ in this model still does not mean complete positive correlation in, e.g., the Pearson correlation coefficient sense, as the connectivity and dependency degrees will not match exactly. An exact match would not be attainable if we want to maintain the possibility of $z_c \neq z_d$.)

To write exact self-consistency equations for the weakly dependent percolation model, we start by writing the degree distribution of the dependency network,

{
\medmuskip=0mu
\thinmuskip=0mu
\thickmuskip=0mu
\begin{align}
P_d(k') = \sum_{k=0}^{\infty} P_c(k)  \left[  (1-f) \frac{z_d^{k'} e^{-z_d}}{k'!} + f \frac{ (k\frac{z_d}{z_c})^{k'} e^{- (k\frac{z_d}{z_c}) }}{k'!}  \right] .
\label{eq6.20}
\end{align}
}

\noindent
Since the connectivity network is uncorrelated, the dependency network is also uncorrelated, with degree distribution given by Eq. (\ref{eq6.20}), and correlations only exist between the connectivity and dependency degrees of nodes.
Let $P(k, k')$ denote the probability that the connectivity degree of a randomly chosen node is $k$ and its dependency degree is $k'$. This can be expressed in two equivalent ways,

\begin{align}
P(k, k') &= P_c(k) \, P(k_d = k' | k_c = k) \nonumber \\
&= P_d(k') \, P(k_c = k | k_d = k').
\label{eq6.30}
\end{align}

\noindent
Using Eqs. (\ref{eq6.10}) and (\ref{eq6.30}) we can write the following,

\begin{align}
&P(k_c = k | k_d = k') = \nonumber \\
&= \frac{P_c(k)}{P_d(k')}  \left[ (1-f) \frac{z_d^{k'} e^{-z_d}}{k'!} + f \frac{ (k\frac{z_d}{z_c})^{k'} e^{- (k\frac{z_d}{z_c}) }}{k'!} \right].
\label{eq6.40}
\end{align}

To proceed it will be useful to express $P_c(k,k')$, the probability that a node arrived at by a random connectivity link, has connectivity degree $k$ and dependency degree $k'$:

\begin{align}
&P_c(k,k') = \nonumber \\
&= \frac{k P_c(k)}{\langle k \rangle_c} P(k_d = k' | k_c = k) \nonumber \\
&= \frac{k P_c(k)}{\langle k \rangle_c} \left[ (1-f) \frac{z_d^{k'} e^{-z_d}}{k'!} + f \frac{ (k\frac{z_d}{z_c})^{k'} e^{- (k\frac{z_d}{z_c}) }}{k'!}  \right].
\label{eq6.50}
\end{align}

\noindent
Similarly, let $P_d(k,k')$ be the probability that a node arrived at by a random dependency link, has connectivity degree $k$ and dependency degree $k'$:

{
\medmuskip=0mu
\thinmuskip=0mu
\thickmuskip=0mu
\begin{align}
&P_d(k,k') = \nonumber \\
&= \frac{k' P_d(k')}{\langle k \rangle_d} P(k_c = k | k_d = k') \nonumber \\
&= \frac{k' P_d(k')}{\langle k \rangle_d} \frac{P_c(k)}{P_d(k')}  \left[ (1-f) \frac{z_d^{k'} e^{-z_d}}{k'!} + f \frac{ (k\frac{z_d}{z_c})^{k'} e^{- (k\frac{z_d}{z_c}) }}{k'!} \right] \nonumber \\
&= \frac{k' P_c(k)}{\langle k \rangle_d}  \left[ (1-f) \frac{z_d^{k'} e^{-z_d}}{k'!} + f \frac{ (k\frac{z_d}{z_c})^{k'} e^{- (k\frac{z_d}{z_c}) }}{k'!}  \right].
\label{eq6.60}
\end{align}
}

\noindent
Finally, let us express the probability $P_{d \to c}(k)$ that a node arrived at by a random dependency link, has connectivity degree $k$,

{
\medmuskip=0mu
\thinmuskip=0mu
\thickmuskip=0mu
\begin{align}
&P_{d \to c}(k) = \nonumber \\
&= \sum_{k'=1}^{\infty} P_d(k,k') \nonumber \\
&= \sum_{k'=1}^{\infty} \frac{k' P_c(k)}{\langle k \rangle_d} \left[  (1-f) \frac{z_d^{k'} e^{-z_d}}{k'!} + f \frac{ (k\frac{z_d}{z_c})^{k'} e^{- (k\frac{z_d}{z_c}) }}{k'!}  \right].
\label{eq6.70}
\end{align}
}

\vskip 0.5cm

With these quantities we can now set up exact self-consistency equations for this correlated model. The probabilities $x$ and $y$ have the same meaning as before.

\begin{align}
x = p \sum_{k=1}^{\infty} \sum_{k'=0}^{\infty}  &\Big\{   P_c(k,k') \left[ 1 - (1-x)^{k-1}  \right] \, \times \\
& \times \, \left[ \delta_{k',0} + 1 - (1-y)^{k'} \right]  \Big\},
\label{eq6.80}
\end{align}

\begin{align}
\label{eq6.90}
y = \sum_{k=0}^{\infty} P_{d \to c}(k) \left[ 1 - (1-x)^k \right].
\end{align}

\noindent
The probability that a random node belongs to the GWDC is now given as

{
\medmuskip=0mu
\thinmuskip=0mu
\thickmuskip=0mu
\begin{align}
S = \sum_{k=0}^{\infty} \sum_{k'=0}^{\infty} P(k,k') \left[ 1 - (1-x)^k  \right] \left[ \delta_{k',0} + 1 - (1-y)^{k'} \right].
\label{eq6.100}
\end{align}
}

\noindent
Plugging Eq. (\ref{eq6.90}) into Eq. (\ref{eq6.80}) results in a single self-consistency equation, and allows for the same kind of analysis as for the uncorrelated model in Sections \ref{sec3} and \ref{sec4}.

\begin{figure}[H]
\centering
\includegraphics[width=\columnwidth,angle=0.]{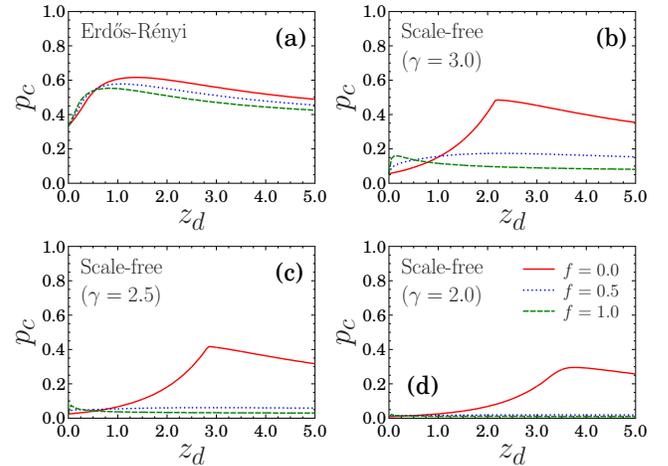}
\caption{Phase diagrams for the model of correlated connectivity and dependency degrees defined by the distribution (\ref{eq6.10}). The percolation threshold values, $p_c$,  were determined by numerical analysis of Eqs. (\ref{eq6.80}) and (\ref{eq6.90}). The mean connectivity degree was $z_c = 3$ for all networks. For scale-free connectivity networks the degree distribution (\ref{eq43.10}) was used, with degree cutoff values $k_{\textrm{min}} = 1$ and $k_{\textrm{max}} = 1000$. [For the dependency degree distributions the cutoff values were changed to $k_{\textrm{min}} = 0$ and $k_{\textrm{max}} = 1200$, because in the model described by Eq. (\ref{eq6.10}) a dependency degree equal to zero is possible and also the biggest dependency degree may be slightly greater than the biggest connectivity degree.]}
\label{fig:correl}
\end{figure}

We considered the above model for different connectivity networks and different degrees of positive correlation between connectivity and dependency degrees. To assess the robustness of these networks, from Eqs. (\ref{eq6.80}) and (\ref{eq6.90}) we numerically identified $p_c$, the lowest link activation probability that provides a nonzero value of $S$. This threshold may correspond to a continuous or a discontinuous transition, but here we are only interested in the value $p_c$ where the transition occurs. Figure \ref{fig:correl} shows $p_c$ as a function of the mean dependency degree, $z_d$, for different values of the correlation parameter $f$. As expected, in Erd\H os-R\'enyi connectivity networks [Fig. \ref{fig:correl} (a)], correlations only have a moderate effect, since the connectivity and dependency degrees in this case cannot be too different, irrespective of the degree of correlation. It is clear, however, that increasing positive correlation increases robustness (i.e., decreases $p_c$) for a relatively dense dependency network. (Note that the opposite must be true for the strong dependency model of \cite{parshani2011critical}.) This can be easily understood qualitatively: higher degree nodes, which are more important for percolation, have more dependency links, i.e., have a higher probability of surviving. Interestingly, the opposite is true for low density dependency networks ($z_d$ smaller than $\approx 0.5$): here higher positive correlation means that high degree nodes have a higher probability of actually having at least one dependency link, as most dependency degrees are now $0$ or $1$. The same dual phenomenon is seen amplified for random scale-free networks: positive correlation dramatically increases robustness for higher density of dependency links, while it decreases robustness for low dependency density [Fig \ref{fig:correl} (b,c,d)]. The effect becomes greater for smaller values of the degree distribution exponent $\gamma$. The low dependency density regime also becomes narrower with decreasing $\gamma$. For maximal positive correlation ($f = 1$) and $\gamma = 2$, the value of $p_c$ is practically the same as the percolation threshold for ordinary percolation 
(without any dependencies), irrespective of the density of dependency links. This suggests that networks, in this model setting, can be efficiently protected from random damage by strong enough positive correlation between connectivity and dependency degrees.
The scale-free networks considered here had a finite degree cutoff, so $p_c$ always remains nonzero, i.e., there is no true ``hyper-resilience'', which is a hallmark of a true, asymptotically power-law degree distribution. It is an interesting problem for future work to check if the percolation threshold can indeed go to zero for the maximally correlated case, for any value of $z_d$, when considering a power-law connectivity degree distribution without a degree cutoff.

\section{Discussion and conclusions}
\label{sec7}

We have investigated the percolation properties of random networks where nodes can have dependency neighbours, but only require that at least one of them be in the same component. Assuming also a random network of dependency links, we have demonstrated that networks in this model are considerably more robust than in the strong dependency model of \cite{parshani2011critical}, due to the redundancy of dependency links for dense enough dependency networks. Our weak dependency model predicts a non-monotonic behaviour of the size of the GWDC as a function of the mean dependency degree: both low and high density of dependency links allows for robust structures, with more fragile structures in between. This also gives a natural scale for ``maximum fragility'' in terms of the mean dependency degree.

Studying Erd\H os-R\'enyi and scale-free connectivity networks we have revealed a wide variety of percolation critical phenomena. Varying the mean dependency degree, continuous and discontinuous hybrid transitions were found for Erd\H os-R\'enyi connectivity networks, separated by a tricritical point. We have derived expressions for the different threshold conditions in terms of the moments of the connectivity and dependency degree distributions. We found standard mean field critical behaviour for continuous transitions, and the same type of critical behaviour for the discontinuous transitions as already seen in $k$-cores and multiplex networks, with order parameter exponent $1/2$. We have identified a special ``cusp'' point, above which the system is always stable, irrespective of the density of dependency links. At this point the order parameter exponent of discontinuous transitions was found to change from $1/2$ to $1$.

We have found continuous and discontinuous hybrid transitions also for scale-free connectivity networks, in the low and high dependency regimes, respectively. For a certain range of the degree distribution exponent $\gamma$ the switch between the two types of transitions was found to be non-smooth, corresponding to the existence of a critical point marking the end of the line of discontinuous transitions. In this range double percolation transitions were observed. The order parameter exponent at the critical point was found to be $1/3$, as seen also in heterogeneous $k$-core, and bootstrap percolation. For the smallest $\gamma$ value where a critical point first appears, we found the exponent to change from $1/3$ to $1/4$ at the (emergent) critical point. The power-law degree distribution used here had a finite cutoff, and hence did not represent a ``truly'' scale-free network. The critical point, double transitions and the unique $1/4$ exponent would likely also appear in other forms of broad degree distributions, not necessarily power-law. These effects can be attributed mainly to the structure of the weak dependency model. Studying the effect of asymptotically power-law degree distributions is an interesting problem for future work.

We have investigated the effect of correlation between connectivity and dependency degrees, and found that positive correlation enhances robustness, except for networks with a low density of dependency links. The robustness enhancing effect is amplified for scale-free connectivity networks. For a low enough value of $\gamma$, strong enough positive correlation between connectivity and dependency degree appears to completely negate the effect of dependencies: the percolation threshold is practically the same as for the network without any dependencies. The study of asymptotically power-law connectivity degree distributions is left for future work.

In addition to degree-degree correlations, network models with a certain fraction of connectivity and dependency links overlapping, may provide a more realistic representation of many real-world systems. Our weak dependency model may also be combined with the strong variant: certain nodes may follow the former rule, others may follow the latter. Such generalizations may be interesting avenues to consider for future research.

\section*{Acknowledgments}

This work was developed within the scope of the project i3N, UIDB/50025/2020 \& UIDP/50025/2020, financed by national funds through the FCT/MEC--Portuguese Foundation for Science and Technology. G. T. was supported by FCT Grant No. CEECIND/03838/2017.


\begin{widetext}

\section*{Appendix: Derivation of order parameter exponents}
\setcounter{equation}{0}
\renewcommand{\theequation}{A\arabic{equation}}

Here we derive the order parameter exponent for the different types of phase transitions that are discussed in the main text. Our starting point is the self-consistency equation

\begin{align}
\label{eqA10}
x \, = \,  p \left[ 1 - H_c(1-x)  \right]   \left[  1 + P_d(0) - G_d(G_c(1-x) )   \right] \, \equiv \, p\Psi(x).
\end{align}

\noindent
In the main text we defined the function $f(x) = \Psi(x) / x$ and discussed the various threshold conditions in terms of the derivatives of $f$. To derive the order parameter exponents it is more convenient to consider the function

\begin{align}
g(x, p) = p \Psi(x) - x. \nonumber
\end{align}

\noindent
The conditions in terms of the derivatives of $f$ can be easily transformed into conditions in terms of the derivatives of $g$. Equation (\ref{eqA10}) can be written as

\begin{align}
g(x, p) = 0. \nonumber
\end{align}

\noindent
Note that $\Psi(0) = 0$, which implies that the ``unmixed'' partial derivatives of $g$ with respect to $p$ are all zero at $x=0$. We will use this fact extensively.

\subsection*{Continuous transitions}
\label{app:cont}

A continuous transition happens at $x=0, p=p_c$, so we expand $g(x,p)$ about the point $(0,p_c)$:

\begin{align}
g(0 + \delta x, p_c + \delta p) &= g(0, p_c) + \frac{\partial g}{\partial x}  \biggr|_{0,p_c} \delta x +
\frac{\partial g}{\partial p}  \biggr|_{0,p_c} \delta p + \frac{1}{2} \frac{\partial^2 g}{\partial x^2}  \biggr|_{0,p_c} (\delta x)^2 + \nonumber \\
&+ \frac{1}{2} \frac{\partial^2 g}{\partial p^2}  \biggr|_{0,p_c} (\delta p)^2 +
\frac{\partial^2 g}{\partial x \partial p}  \biggr|_{0,p_c} (\delta x)(\delta p) + \ldots.
\label{eqA30}
\end{align}

\noindent
We know that $g(0, p_c) = 0$. Requiring also that $g(0 + \delta x, p_c + \delta p) = 0$ (i.e., we are expanding along the solution curve) we have that the expansion terms of equal powers on the right-hand side of Eq. (\ref{eqA30}) must cancel out.
The unmixed partial derivatives of $g$ with respect to $p$ are all $0$ at $x=0$. The condition for a continuous transition, as expressed in terms of the function $f$, is

\begin{align}
p_c \left( \lim_{x \to 0} f(x) \right) = 1. \nonumber
\end{align}

\noindent
This condition is equivalent to

\begin{align}
g(0, p_c) = \frac{\partial g}{\partial x}  \biggr|_{0,p_c} = 0. \nonumber
\end{align}

\noindent
With the above conditions, the leading terms of Eq. (\ref{eqA30}) are

\begin{align}
\frac{1}{2} \frac{\partial^2 g}{\partial x^2}  \biggr|_{0,p_c} (\delta x)^2 +
\frac{\partial^2 g}{\partial x \partial p}  \biggr|_{0,p_c} (\delta x)(\delta p)  = 0, \nonumber
\end{align}

\noindent
which results in

\begin{align}
\delta x \propto \delta p. \nonumber
\end{align}

\subsection*{Discontinuous transitions}
\label{app:disc}

A discontinuous transition happens at some $x=x^*, p=p_c$, so we expand $g(x,p)$ about the point $(x^*,p_c)$:

\begin{align}
g(x^* + \delta x, p_c + \delta p) &= g(x^*, p_c) + \frac{\partial g}{\partial x}  \biggr|_{x^*,p_c} \delta x +
\frac{\partial g}{\partial p}  \biggr|_{x^*,p_c} \delta p + \frac{1}{2} \frac{\partial^2 g}{\partial x^2}  \biggr|_{x^*,p_c} (\delta x)^2 + \nonumber \\
&+ \frac{1}{2} \frac{\partial^2 g}{\partial p^2}  \biggr|_{x^*,p_c} (\delta p)^2 +
\frac{\partial^2 g}{\partial x \partial p}  \biggr|_{x^*,p_c} (\delta x)(\delta p) + \ldots.
\label{eqA90}
\end{align}

\noindent
In this case, since $x^* > 0$, the unmixed derivatives of $g$ with respect to $p$ are nonzero.
The conditions for a discontinuous transition, as expressed in terms of the function $f$, are

\begin{align}
p_c f(x^*) &= 1, \nonumber \\
f'(x^*) &= 0. \nonumber
\end{align}

\noindent
These conditions are equivalent to

\begin{align}
g(x^*, p_c) = \frac{\partial g}{\partial x}  \biggr|_{x^*,p_c} = 0. \nonumber
\end{align}

\noindent
With the above conditions, the leading terms in Eq. (\ref{eqA90}) are

\begin{align}
\frac{\partial g}{\partial p}  \biggr|_{x^*, p_c} \delta p + \frac{1}{2} \frac{\partial^2 g}{\partial x^2}  \biggr|_{x^*, p_c} (\delta x)^2 = 0, \nonumber
\end{align}

\noindent
which results in

\begin{align}
\delta x \propto (\delta p)^{1/2}. \nonumber
\end{align}

\subsection*{Tricritical point}
\label{app:tric}

A tricritical point occurs at $x=0, p=p_c$, so we expand $g(x,p)$ about the point $(0,p_c)$:

\begin{align}
g(0 + \delta x, p_c + \delta p) &= g(0, p_c) + \frac{\partial g}{\partial x}  \biggr|_{0,p_c} \delta x +
\frac{\partial g}{\partial p}  \biggr|_{0,p_c} \delta p + \frac{1}{2} \frac{\partial^2 g}{\partial x^2}  \biggr|_{0,p_c} (\delta x)^2 + \nonumber \\
&+ \frac{1}{2} \frac{\partial^2 g}{\partial p^2}  \biggr|_{0,p_c} (\delta p)^2 +
\frac{\partial^2 g}{\partial x \partial p}  \biggr|_{0,p_c} (\delta x)(\delta p) + \nonumber \\
&+ \frac{1}{6} \frac{\partial^3 g}{\partial x^3}  \biggr|_{0, p_c} (\delta x)^3 +
\frac{1}{6} \frac{\partial^3 g}{\partial p^3}  \biggr|_{0, p_c} (\delta p)^3 + \nonumber \\
&+ \frac{1}{2} \frac{\partial^3 g}{\partial x^2 \partial p}  \biggr|_{0, p_c} (\delta x)^2 (\delta p) +
\frac{1}{2} \frac{\partial^3 g}{\partial x \partial p^2}  \biggr|_{0, p_c} (\delta x) (\delta p)^2 + \ldots.
\label{eqA140}
\end{align}

\noindent
As before, for continuous transitions, the unmixed derivatives of $g$ with respect to $p$ are all $0$.
The conditions for a tricritical point, as expressed in terms of the function $f$, are

\begin{align}
p_c \left( \lim_{x \to 0} f(x) \right) &= 1, \nonumber \\
\lim_{x \to 0} f'(x) &= 0. \nonumber
\end{align}

\noindent
These conditions are equivalent to

\begin{align}
g(0, p_c) = \frac{\partial g}{\partial x}  \biggr|_{0,p_c} = \frac{\partial^2 g}{\partial x^2}  \biggr|_{0,p_c} = 0. \nonumber
\end{align}

\noindent
With the above conditions, the leading terms in Eq. (\ref{eqA140}) are

\begin{align}
&\frac{\partial^2 g}{\partial x \partial p}  \biggr|_{0, p_c} (\delta x)(\delta p) + \frac{1}{6} \frac{\partial^3 g}{\partial x^3}  \biggr|_{0, p_c} (\delta x)^3 = 0, \nonumber
\end{align}

\noindent
which results in

\begin{align}
\delta x \propto (\delta p)^{1/2}. \nonumber
\end{align}

\subsection*{Cusp point}
\label{app:cusp}

To describe this point, as discussed in the main text, we assume $p = 1$ and we consider $f$ to be a function of $x$ and $z_d$ (the mean dependency degree). The function $g$ is now also assumed to be a function of $x$ and $z_d$: $g(x, z_d) = \Psi(x, z_d) - x = x f(x,z_d) - x$. The cusp point occurs at some $x=x^{\textrm{cp}}, z_d = z_d^{\textrm{cp}}$, so we expand $g(x,z_d)$ about the point $(x^{\textrm{cp}},z_d^{\textrm{cp}})$:

\begin{align}
g(x^{\textrm{cp}} + \delta x, z_d^{\textrm{cp}} + \delta z_d) &= g(x^{\textrm{cp}}, z_d^{\textrm{cp}}) + \frac{\partial g}{\partial x}  \biggr|_{x^{\textrm{cp}},z_d^{\textrm{cp}}} \delta x +
\frac{\partial g}{\partial z_d}  \biggr|_{x^{\textrm{cp}},z_d^{\textrm{cp}}} \delta z_d + \frac{1}{2} \frac{\partial^2 g}{\partial x^2}  \biggr|_{x^{\textrm{cp}},z_d^{\textrm{cp}}} (\delta x)^2 + \nonumber \\
&+ \frac{1}{2} \frac{\partial^2 g}{\partial z_d^2}  \biggr|_{x^{\textrm{cp}},z_d^{\textrm{cp}}} (\delta z_d)^2 +
\frac{\partial^2 g}{\partial x \partial z_d}  \biggr|_{x^{\textrm{cp}},z_d^{\textrm{cp}}} (\delta x)(\delta z_d) + \ldots.
\label{eqA181}
\end{align}

\noindent
The conditions for the cusp point, as expressed in terms of the function $f$, are

\begin{align}
f(x^{\textrm{cp}}, z_d^{\textrm{cp}}) &= 1, \nonumber \\
\frac{\partial f}{\partial x}  \biggr|_{x^{\textrm{cp}},z_d^{\textrm{cp}}} &= 0 \nonumber \\
\frac{\partial f}{\partial z_d}  \biggr|_{x^{\textrm{cp}},z_d^{\textrm{cp}}} &= 0. \nonumber
\end{align}

\noindent
These conditions are equivalent to 

\begin{align}
g(x^{\textrm{cp}}, z_d^{\textrm{cp}}) = 
\frac{\partial g}{\partial x}  \biggr|_{x^{\textrm{cp}},z_d^{\textrm{cp}}} = 
\frac{\partial g}{\partial z_d}  \biggr|_{x^{\textrm{cp}},z_d^{\textrm{cp}}} = 0.  \nonumber
\end{align}

\noindent
With the above conditions, the leading terms in Eq. (\ref{eqA181}) are

\begin{align}
\frac{1}{2} \frac{\partial^2 g}{\partial x^2}  \biggr|_{x^{\textrm{cp}},z_d^{\textrm{cp}}} (\delta x)^2 +
\frac{1}{2} \frac{\partial^2 g}{\partial z_d^2}  \biggr|_{x^{\textrm{cp}},z_d^{\textrm{cp}}} (\delta z_d)^2 +
\frac{\partial^2 g}{\partial x \partial z_d}  \biggr|_{x^{\textrm{cp}},z_d^{\textrm{cp}}} (\delta x)(\delta z_d) = 0,  \nonumber
\end{align}

\noindent
which results in

\begin{align}
\delta x \propto \delta z_d. \nonumber
\end{align}


\subsection*{Critical point}
\label{app:crit}

A critical point occurs at some $x=x^*, p=p_c$, so we expand $g(x,p)$ about the point $(x^*,p_c)$:

\begin{align}
g(x^* + \delta x, p_c + \delta p) &= g(x^*, p_c) + \frac{\partial g}{\partial x}  \biggr|_{x^*,p_c} \delta x +
\frac{\partial g}{\partial p}  \biggr|_{x^*,p_c} \delta p + \frac{1}{2} \frac{\partial^2 g}{\partial x^2}  \biggr|_{x^*,p_c} (\delta x)^2 + \nonumber \\
&+ \frac{1}{2} \frac{\partial^2 g}{\partial p^2}  \biggr|_{x^*,p_c} (\delta p)^2 +
\frac{\partial^2 g}{\partial x \partial p}  \biggr|_{x^*,p_c} (\delta x)(\delta p) + \nonumber \\
&+ \frac{1}{6} \frac{\partial^3 g}{\partial x^3}  \biggr|_{x^*, p_c} (\delta x)^3 +
\frac{1}{6} \frac{\partial^3 g}{\partial p^3}  \biggr|_{x^*, p_c} (\delta p)^3 + \nonumber \\
&+ \frac{1}{2} \frac{\partial^3 g}{\partial x^2 \partial p}  \biggr|_{x^*, p_c} (\delta x)^2 (\delta p) +
\frac{1}{2} \frac{\partial^3 g}{\partial x \partial p^2}  \biggr|_{x^*, p_c} (\delta x) (\delta p)^2 + \ldots.
\label{eqA190}
\end{align}

\noindent
Here, as in the case of discontinuous transitions, the unmixed derivatives of $g$ with respect to $p$ are nonzero.
The conditions for a critical point, as expressed in terms of the function $f$, are

\begin{align}
p_c f(x^*) &= 1, \nonumber \\
f'(x^*) &= 0, \nonumber \\
f''(x^*) &= 0. \nonumber
\end{align}

\noindent
These conditions are equivalent to

\begin{align}
g(x^*, p_c) = \frac{\partial g}{\partial x}  \biggr|_{x^*,p_c} = \frac{\partial^2 g}{\partial x^2}  \biggr|_{x^*,p_c} = 0. \nonumber
\end{align}

\noindent
With the above conditions, the leading terms in Eq. (\ref{eqA190}) are

\begin{align}
&\frac{\partial g}{\partial p}  \biggr|_{x^*, p_c} \delta p + \frac{1}{6} \frac{\partial^3 g}{\partial x^3}  \biggr|_{x^*, p_c} (\delta x)^3 = 0, \nonumber
\end{align}

\noindent
which results in

\begin{align}
\delta x \propto (\delta p)^{1/3}. \nonumber
\end{align}


\subsection*{Birth point of critical point (at $\gamma^{\textrm{(high)}}$)}
\label{app:high}

The critical point, at $\gamma^{\textrm{(high)}}$ emerges from a tricritical point, which always occurs at $x=0, p=p_c$, so we expand $g(x,p)$ about the point $(0,p_c)$:

\begin{align}
g(0 + \delta x, p_c + \delta p) &= g(0, p_c) + \frac{\partial g}{\partial x}  \biggr|_{0,p_c} \delta x +
\frac{\partial g}{\partial p}  \biggr|_{0,p_c} \delta p + \frac{1}{2} \frac{\partial^2 g}{\partial x^2}  \biggr|_{0,p_c} (\delta x)^2 + \nonumber \\
&+ \frac{1}{2} \frac{\partial^2 g}{\partial p^2}  \biggr|_{0,p_c} (\delta p)^2 +
\frac{\partial^2 g}{\partial x \partial p}  \biggr|_{0,p_c} (\delta x)(\delta p) + \nonumber \\
&+ \frac{1}{6} \frac{\partial^3 g}{\partial x^3}  \biggr|_{0, p_c} (\delta x)^3 +
\frac{1}{6} \frac{\partial^3 g}{\partial p^3}  \biggr|_{0, p_c} (\delta p)^3 + \nonumber \\
&+ \frac{1}{2} \frac{\partial^3 g}{\partial x^2 \partial p}  \biggr|_{0, p_c} (\delta x)^2 (\delta p) +
\frac{1}{2} \frac{\partial^3 g}{\partial x \partial p^2}  \biggr|_{0, p_c} (\delta x) (\delta p)^2 + \ldots.
\label{eqA260}
\end{align}

\noindent
The unmixed derivatives of $g$ with respect to $p$ are $0$.
The conditions for the birth point of the critical point at $x=0$, as expressed in terms of the function $f$, are

\begin{align}
p_c \left(  \lim_{x \to 0} f(x) \right) &= 1, \nonumber \\
\lim_{x \to 0} f'(x) &= 0, \nonumber \\
\lim_{x \to 0} f''(x) &= 0. \nonumber
\end{align}

\noindent
These conditions are equivalent to

\begin{align}
g(0, p_c) = \frac{\partial g}{\partial x}  \biggr|_{0,p_c} = \frac{\partial^2 g}{\partial x^2}  \biggr|_{0,p_c} = \frac{\partial^3 g}{\partial x^3}  \biggr|_{0,p_c} = 0. \nonumber
\end{align}

\noindent
With the above conditions, the leading terms in Eq. (\ref{eqA260}) are

\begin{align}
&\frac{\partial^2 g}{\partial x \partial p}  \biggr|_{0, p_c} (\delta x)(\delta p) + \frac{1}{24} \frac{\partial^4 g}{\partial x^4}  \biggr|_{0, p_c} (\delta x)^4 = 0, \nonumber
\end{align}

\noindent
which results in

\begin{align}
\delta x \propto (\delta p)^{1/3}. \nonumber
\end{align}


\subsection*{Birth point of critical point (at $\gamma^{\textrm{(low)}}$)}
\label{app:low}

The critical point, at $\gamma^{\textrm{(low)}}$, emerges at some $x=x^*, p=p_c$, so we expand $g(x,p)$ about the point $(x^*,p_c)$:

\begin{align}
g(x^* + \delta x, p_c + \delta p) &= g(x^*, p_c) + \frac{\partial g}{\partial x}  \biggr|_{x^*,p_c} \delta x +
\frac{\partial g}{\partial p}  \biggr|_{x^*,p_c} \delta p + \frac{1}{2} \frac{\partial^2 g}{\partial x^2}  \biggr|_{x^*,p_c} (\delta x)^2 + \nonumber \\
&+ \frac{1}{2} \frac{\partial^2 g}{\partial p^2}  \biggr|_{x^*,p_c} (\delta p)^2 +
\frac{\partial^2 g}{\partial x \partial p}  \biggr|_{x^*,p_c} (\delta x)(\delta p) + \nonumber \\
&+ \frac{1}{6} \frac{\partial^3 g}{\partial x^3}  \biggr|_{x^*, p_c} (\delta x)^3 +
\frac{1}{6} \frac{\partial^3 g}{\partial p^3}  \biggr|_{x^*, p_c} (\delta p)^3 + \nonumber \\
&+ \frac{1}{2} \frac{\partial^3 g}{\partial x^2 \partial p}  \biggr|_{x^*, p_c} (\delta x)^2 (\delta p) +
\frac{1}{2} \frac{\partial^3 g}{\partial x \partial p^2}  \biggr|_{x^*, p_c} (\delta x) (\delta p)^2 + \ldots.
\label{eqA330}
\end{align}

\noindent
The unmixed derivatives of $g$ with respect to $p$ are nonzero.
The conditions for the birth point of the critical point, as expressed in terms of the function $f$, are

\begin{align}
p_c f(x^*) &= 1, \nonumber \\
f'(x^*) &= 0, \nonumber \\
f''(x^*) &= 0, \nonumber \\
f'''(x^*) &= 0. \nonumber \\
\end{align}

\noindent
These conditions are equivalent to

\begin{align}
g(x^*, p_c) = \frac{\partial g}{\partial x}  \biggr|_{x^*,p_c} = \frac{\partial^2 g}{\partial x^2}  \biggr|_{x^*,p_c} = \frac{\partial^3 g}{\partial x^3}  \biggr|_{x^*,p_c} = 0. \nonumber
\end{align}

\noindent
With the above conditions, the leading terms in Eq. (\ref{eqA330}) are

\begin{align}
&\frac{\partial g}{\partial p}  \biggr|_{x^*, p_c} \delta p + \frac{1}{24} \frac{\partial^4 g}{\partial x^4}  \biggr|_{x^*, p_c} (\delta x)^4 = 0, \nonumber
\end{align}

\noindent
which results in

\begin{align}
\delta x \propto (\delta p)^{1/4}. \nonumber
\end{align}

\end{widetext}



\begin{thebibliography}{23}%
\makeatletter
\providecommand \@ifxundefined [1]{%
 \@ifx{#1\undefined}
}%
\providecommand \@ifnum [1]{%
 \ifnum #1\expandafter \@firstoftwo
 \else \expandafter \@secondoftwo
 \fi
}%
\providecommand \@ifx [1]{%
 \ifx #1\expandafter \@firstoftwo
 \else \expandafter \@secondoftwo
 \fi
}%
\providecommand \natexlab [1]{#1}%
\providecommand \enquote  [1]{``#1''}%
\providecommand \bibnamefont  [1]{#1}%
\providecommand \bibfnamefont [1]{#1}%
\providecommand \citenamefont [1]{#1}%
\providecommand \href@noop [0]{\@secondoftwo}%
\providecommand \href [0]{\begingroup \@sanitize@url \@href}%
\providecommand \@href[1]{\@@startlink{#1}\@@href}%
\providecommand \@@href[1]{\endgroup#1\@@endlink}%
\providecommand \@sanitize@url [0]{\catcode `\\12\catcode `\$12\catcode
  `\&12\catcode `\#12\catcode `\^12\catcode `\_12\catcode `\%12\relax}%
\providecommand \@@startlink[1]{}%
\providecommand \@@endlink[0]{}%
\providecommand \url  [0]{\begingroup\@sanitize@url \@url }%
\providecommand \@url [1]{\endgroup\@href {#1}{\urlprefix }}%
\providecommand \urlprefix  [0]{URL }%
\providecommand \Eprint [0]{\href }%
\providecommand \doibase [0]{http://dx.doi.org/}%
\providecommand \selectlanguage [0]{\@gobble}%
\providecommand \bibinfo  [0]{\@secondoftwo}%
\providecommand \bibfield  [0]{\@secondoftwo}%
\providecommand \translation [1]{[#1]}%
\providecommand \BibitemOpen [0]{}%
\providecommand \bibitemStop [0]{}%
\providecommand \bibitemNoStop [0]{.\EOS\space}%
\providecommand \EOS [0]{\spacefactor3000\relax}%
\providecommand \BibitemShut  [1]{\csname bibitem#1\endcsname}%
\let\auto@bib@innerbib\@empty
\bibitem [{\citenamefont {Buldyrev}\ \emph {et~al.}(2010)\citenamefont
  {Buldyrev}, \citenamefont {Parshani}, \citenamefont {Paul}, \citenamefont
  {Stanley},\ and\ \citenamefont {Havlin}}]{buldyrev2010catastrophic}%
  \BibitemOpen
  \bibfield  {author} {\bibinfo {author} {\bibfnamefont {S.~V.}\ \bibnamefont
  {Buldyrev}}, \bibinfo {author} {\bibfnamefont {R.}~\bibnamefont {Parshani}},
  \bibinfo {author} {\bibfnamefont {G.}~\bibnamefont {Paul}}, \bibinfo {author}
  {\bibfnamefont {H.~E.}\ \bibnamefont {Stanley}}, \ and\ \bibinfo {author}
  {\bibfnamefont {S.}~\bibnamefont {Havlin}},\ }\bibfield  {title} {\enquote
  {\bibinfo {title} {Catastrophic cascade of failures in interdependent
  networks},}\ }\href@noop {} {\bibfield  {journal} {\bibinfo  {journal}
  {Nature}\ }\textbf {\bibinfo {volume} {464}},\ \bibinfo {pages} {1025}
  (\bibinfo {year} {2010})}\BibitemShut {NoStop}%
\bibitem [{\citenamefont {Son}\ \emph {et~al.}(2012)\citenamefont {Son},
  \citenamefont {Bizhani}, \citenamefont {Christensen}, \citenamefont
  {Grassberger},\ and\ \citenamefont {Paczuski}}]{son2012percolation}%
  \BibitemOpen
  \bibfield  {author} {\bibinfo {author} {\bibfnamefont {S.-W.}\ \bibnamefont
  {Son}}, \bibinfo {author} {\bibfnamefont {G.}~\bibnamefont {Bizhani}},
  \bibinfo {author} {\bibfnamefont {C.}~\bibnamefont {Christensen}}, \bibinfo
  {author} {\bibfnamefont {P.}~\bibnamefont {Grassberger}}, \ and\ \bibinfo
  {author} {\bibfnamefont {M.}~\bibnamefont {Paczuski}},\ }\bibfield  {title}
  {\enquote {\bibinfo {title} {Percolation theory on interdependent networks
  based on epidemic spreading},}\ }\href@noop {} {\bibfield  {journal}
  {\bibinfo  {journal} {Europhys. Lett.}\ }\textbf {\bibinfo {volume} {97}},\
  \bibinfo {pages} {16006} (\bibinfo {year} {2012})}\BibitemShut {NoStop}%
\bibitem [{\citenamefont {Baxter}\ \emph {et~al.}(2012)\citenamefont {Baxter},
  \citenamefont {Dorogovtsev}, \citenamefont {Goltsev},\ and\ \citenamefont
  {Mendes}}]{baxter2012avalanche}%
  \BibitemOpen
  \bibfield  {author} {\bibinfo {author} {\bibfnamefont {G.~J.}\ \bibnamefont
  {Baxter}}, \bibinfo {author} {\bibfnamefont {S.~N.}\ \bibnamefont
  {Dorogovtsev}}, \bibinfo {author} {\bibfnamefont {A.~V.}\ \bibnamefont
  {Goltsev}}, \ and\ \bibinfo {author} {\bibfnamefont {J.~F.~F.}\ \bibnamefont
  {Mendes}},\ }\bibfield  {title} {\enquote {\bibinfo {title} {Avalanche
  collapse of interdependent networks},}\ }\href@noop {} {\bibfield  {journal}
  {\bibinfo  {journal} {Phys. Rev. Lett.}\ }\textbf {\bibinfo {volume} {109}},\
  \bibinfo {pages} {248701} (\bibinfo {year} {2012})}\BibitemShut {NoStop}%
\bibitem [{\citenamefont {Dorogovtsev}\ \emph {et~al.}(2006)\citenamefont
  {Dorogovtsev}, \citenamefont {Goltsev},\ and\ \citenamefont
  {Mendes}}]{dorogovtsev2006k}%
  \BibitemOpen
  \bibfield  {author} {\bibinfo {author} {\bibfnamefont {S.~N.}\ \bibnamefont
  {Dorogovtsev}}, \bibinfo {author} {\bibfnamefont {A.~V.}\ \bibnamefont
  {Goltsev}}, \ and\ \bibinfo {author} {\bibfnamefont {J.~F.~F.}\ \bibnamefont
  {Mendes}},\ }\bibfield  {title} {\enquote {\bibinfo {title} {K-core
  organization of complex networks},}\ }\href@noop {} {\bibfield  {journal}
  {\bibinfo  {journal} {Phys. Rev. Lett.}\ }\textbf {\bibinfo {volume} {96}},\
  \bibinfo {pages} {040601} (\bibinfo {year} {2006})}\BibitemShut {NoStop}%
\bibitem [{\citenamefont {Boccaletti}\ \emph {et~al.}(2014)\citenamefont
  {Boccaletti}, \citenamefont {Bianconi}, \citenamefont {Criado}, \citenamefont
  {Del~Genio}, \citenamefont {G{\'o}mez-Gardenes}, \citenamefont {Romance},
  \citenamefont {Sendina-Nadal}, \citenamefont {Wang},\ and\ \citenamefont
  {Zanin}}]{boccaletti2014structure}%
  \BibitemOpen
  \bibfield  {author} {\bibinfo {author} {\bibfnamefont {S.}~\bibnamefont
  {Boccaletti}}, \bibinfo {author} {\bibfnamefont {G.}~\bibnamefont
  {Bianconi}}, \bibinfo {author} {\bibfnamefont {R.}~\bibnamefont {Criado}},
  \bibinfo {author} {\bibfnamefont {C.~I.}\ \bibnamefont {Del~Genio}}, \bibinfo
  {author} {\bibfnamefont {J.}~\bibnamefont {G{\'o}mez-Gardenes}}, \bibinfo
  {author} {\bibfnamefont {M.}~\bibnamefont {Romance}}, \bibinfo {author}
  {\bibfnamefont {I.}~\bibnamefont {Sendina-Nadal}}, \bibinfo {author}
  {\bibfnamefont {Z.}~\bibnamefont {Wang}}, \ and\ \bibinfo {author}
  {\bibfnamefont {M.}~\bibnamefont {Zanin}},\ }\bibfield  {title} {\enquote
  {\bibinfo {title} {The structure and dynamics of multilayer networks},}\
  }\href@noop {} {\bibfield  {journal} {\bibinfo  {journal} {Physics Reports}\
  }\textbf {\bibinfo {volume} {544}},\ \bibinfo {pages} {1} (\bibinfo {year}
  {2014})}\BibitemShut {NoStop}%
\bibitem [{\citenamefont {Parshani}\ \emph {et~al.}(2011)\citenamefont
  {Parshani}, \citenamefont {Buldyrev},\ and\ \citenamefont
  {Havlin}}]{parshani2011critical}%
  \BibitemOpen
  \bibfield  {author} {\bibinfo {author} {\bibfnamefont {R.}~\bibnamefont
  {Parshani}}, \bibinfo {author} {\bibfnamefont {S.~V.}\ \bibnamefont
  {Buldyrev}}, \ and\ \bibinfo {author} {\bibfnamefont {S.}~\bibnamefont
  {Havlin}},\ }\bibfield  {title} {\enquote {\bibinfo {title} {Critical effect
  of dependency groups on the function of networks},}\ }\href@noop {}
  {\bibfield  {journal} {\bibinfo  {journal} {PNAS}\ }\textbf {\bibinfo
  {volume} {108}},\ \bibinfo {pages} {1007} (\bibinfo {year}
  {2011})}\BibitemShut {NoStop}%
\bibitem [{\citenamefont {Bashan}\ \emph {et~al.}(2011)\citenamefont {Bashan},
  \citenamefont {Parshani},\ and\ \citenamefont
  {Havlin}}]{bashan2011percolation}%
  \BibitemOpen
  \bibfield  {author} {\bibinfo {author} {\bibfnamefont {A.}~\bibnamefont
  {Bashan}}, \bibinfo {author} {\bibfnamefont {R.}~\bibnamefont {Parshani}}, \
  and\ \bibinfo {author} {\bibfnamefont {S.}~\bibnamefont {Havlin}},\
  }\bibfield  {title} {\enquote {\bibinfo {title} {Percolation in networks
  composed of connectivity and dependency links},}\ }\href@noop {} {\bibfield
  {journal} {\bibinfo  {journal} {Phys. Rev. E}\ }\textbf {\bibinfo {volume}
  {83}},\ \bibinfo {pages} {051127} (\bibinfo {year} {2011})}\BibitemShut
  {NoStop}%
\bibitem [{\citenamefont {Bashan}\ and\ \citenamefont
  {Havlin}(2011)}]{bashan2011combined}%
  \BibitemOpen
  \bibfield  {author} {\bibinfo {author} {\bibfnamefont {A.}~\bibnamefont
  {Bashan}}\ and\ \bibinfo {author} {\bibfnamefont {S.}~\bibnamefont
  {Havlin}},\ }\bibfield  {title} {\enquote {\bibinfo {title} {The combined
  effect of connectivity and dependency links on percolation of networks},}\
  }\href@noop {} {\bibfield  {journal} {\bibinfo  {journal} {J. Stat. Phys.}\
  }\textbf {\bibinfo {volume} {145}},\ \bibinfo {pages} {686} (\bibinfo {year}
  {2011})}\BibitemShut {NoStop}%
\bibitem [{\citenamefont {Lin}\ \emph {et~al.}(2017)\citenamefont {Lin},
  \citenamefont {Kang}, \citenamefont {Wang}, \citenamefont {Zhao},
  \citenamefont {Li},\ and\ \citenamefont {Havlin}}]{lin2017robustness}%
  \BibitemOpen
  \bibfield  {author} {\bibinfo {author} {\bibfnamefont {Y.}~\bibnamefont
  {Lin}}, \bibinfo {author} {\bibfnamefont {R.}~\bibnamefont {Kang}}, \bibinfo
  {author} {\bibfnamefont {Z.}~\bibnamefont {Wang}}, \bibinfo {author}
  {\bibfnamefont {Z.}~\bibnamefont {Zhao}}, \bibinfo {author} {\bibfnamefont
  {D.}~\bibnamefont {Li}}, \ and\ \bibinfo {author} {\bibfnamefont
  {S.}~\bibnamefont {Havlin}},\ }\bibfield  {title} {\enquote {\bibinfo {title}
  {Robustness of networks with dependency topology},}\ }\href@noop {}
  {\bibfield  {journal} {\bibinfo  {journal} {Europhys. Lett.}\ }\textbf
  {\bibinfo {volume} {118}},\ \bibinfo {pages} {36002} (\bibinfo {year}
  {2017})}\BibitemShut {NoStop}%
\bibitem [{\citenamefont {Niu}\ \emph {et~al.}(2016)\citenamefont {Niu},
  \citenamefont {Yuan}, \citenamefont {Du}, \citenamefont {Stanley},\ and\
  \citenamefont {Hu}}]{niu2016percolation}%
  \BibitemOpen
  \bibfield  {author} {\bibinfo {author} {\bibfnamefont {D.}~\bibnamefont
  {Niu}}, \bibinfo {author} {\bibfnamefont {X.}~\bibnamefont {Yuan}}, \bibinfo
  {author} {\bibfnamefont {M.}~\bibnamefont {Du}}, \bibinfo {author}
  {\bibfnamefont {H.~E.}\ \bibnamefont {Stanley}}, \ and\ \bibinfo {author}
  {\bibfnamefont {Y.}~\bibnamefont {Hu}},\ }\bibfield  {title} {\enquote
  {\bibinfo {title} {Percolation of networks with directed dependency links},}\
  }\href@noop {} {\bibfield  {journal} {\bibinfo  {journal} {Phys. Rev. E}\
  }\textbf {\bibinfo {volume} {93}},\ \bibinfo {pages} {042312} (\bibinfo
  {year} {2016})}\BibitemShut {NoStop}%
\bibitem [{\citenamefont {Bai}\ \emph {et~al.}(2016)\citenamefont {Bai},
  \citenamefont {Huang}, \citenamefont {Wang},\ and\ \citenamefont
  {Wu}}]{bai2016robustness}%
  \BibitemOpen
  \bibfield  {author} {\bibinfo {author} {\bibfnamefont {Y.-N.}\ \bibnamefont
  {Bai}}, \bibinfo {author} {\bibfnamefont {N.}~\bibnamefont {Huang}}, \bibinfo
  {author} {\bibfnamefont {L.}~\bibnamefont {Wang}}, \ and\ \bibinfo {author}
  {\bibfnamefont {Z.-X.}\ \bibnamefont {Wu}},\ }\bibfield  {title} {\enquote
  {\bibinfo {title} {Robustness and vulnerability of networks with dynamical
  dependency groups},}\ }\href@noop {} {\bibfield  {journal} {\bibinfo
  {journal} {Sci. Rep.}\ }\textbf {\bibinfo {volume} {6}},\ \bibinfo {pages}
  {37749} (\bibinfo {year} {2016})}\BibitemShut {NoStop}%
\bibitem [{\citenamefont {Liu}\ \emph {et~al.}(2016)\citenamefont {Liu},
  \citenamefont {Li}, \citenamefont {Jia},\ and\ \citenamefont
  {Wang}}]{liu2016cascading}%
  \BibitemOpen
  \bibfield  {author} {\bibinfo {author} {\bibfnamefont {R.-R.}\ \bibnamefont
  {Liu}}, \bibinfo {author} {\bibfnamefont {M.}~\bibnamefont {Li}}, \bibinfo
  {author} {\bibfnamefont {C.-X.}\ \bibnamefont {Jia}}, \ and\ \bibinfo
  {author} {\bibfnamefont {B.-H.}\ \bibnamefont {Wang}},\ }\bibfield  {title}
  {\enquote {\bibinfo {title} {Cascading failures in coupled networks with both
  inner-dependency and inter-dependency links},}\ }\href@noop {} {\bibfield
  {journal} {\bibinfo  {journal} {Sci. Rep.}\ }\textbf {\bibinfo {volume}
  {6}},\ \bibinfo {pages} {25294} (\bibinfo {year} {2016})}\BibitemShut
  {NoStop}%
\bibitem [{\citenamefont {Baxter}\ \emph {et~al.}(2014)\citenamefont {Baxter},
  \citenamefont {Dorogovtsev}, \citenamefont {Mendes},\ and\ \citenamefont
  {Cellai}}]{baxter2014weak}%
  \BibitemOpen
  \bibfield  {author} {\bibinfo {author} {\bibfnamefont {G.~J.}\ \bibnamefont
  {Baxter}}, \bibinfo {author} {\bibfnamefont {S.~N.}\ \bibnamefont
  {Dorogovtsev}}, \bibinfo {author} {\bibfnamefont {J.~F.~F.}\ \bibnamefont
  {Mendes}}, \ and\ \bibinfo {author} {\bibfnamefont {D.}~\bibnamefont
  {Cellai}},\ }\bibfield  {title} {\enquote {\bibinfo {title} {Weak percolation
  on multiplex networks},}\ }\href@noop {} {\bibfield  {journal} {\bibinfo
  {journal} {Phys. Rev. E}\ }\textbf {\bibinfo {volume} {89}},\ \bibinfo
  {pages} {042801} (\bibinfo {year} {2014})}\BibitemShut {NoStop}%
\bibitem [{\citenamefont {Baxter}\ \emph {et~al.}(2020)\citenamefont {Baxter},
  \citenamefont {da~Costa}, \citenamefont {Dorogovtsev},\ and\ \citenamefont
  {Mendes}}]{baxter2020exotic}%
  \BibitemOpen
  \bibfield  {author} {\bibinfo {author} {\bibfnamefont {G.~J.}\ \bibnamefont
  {Baxter}}, \bibinfo {author} {\bibfnamefont {R.~A.}\ \bibnamefont
  {da~Costa}}, \bibinfo {author} {\bibfnamefont {S.~N.}\ \bibnamefont
  {Dorogovtsev}}, \ and\ \bibinfo {author} {\bibfnamefont {J.~F.~F.}\
  \bibnamefont {Mendes}},\ }\bibfield  {title} {\enquote {\bibinfo {title}
  {Exotic critical behavior of weak multiplex percolation},}\ }\href@noop {}
  {\bibfield  {journal} {\bibinfo  {journal} {Phys. Rev. E}\ }\textbf {\bibinfo
  {volume} {102}},\ \bibinfo {pages} {032301} (\bibinfo {year}
  {2020})}\BibitemShut {NoStop}%
\bibitem [{\citenamefont {Radicchi}\ and\ \citenamefont
  {Bianconi}(2017)}]{radicchi2017redundant}%
  \BibitemOpen
  \bibfield  {author} {\bibinfo {author} {\bibfnamefont {F.}~\bibnamefont
  {Radicchi}}\ and\ \bibinfo {author} {\bibfnamefont {G.}~\bibnamefont
  {Bianconi}},\ }\bibfield  {title} {\enquote {\bibinfo {title} {Redundant
  interdependencies boost the robustness of multiplex networks},}\ }\href@noop
  {} {\bibfield  {journal} {\bibinfo  {journal} {Phys. Rev. X}\ }\textbf
  {\bibinfo {volume} {7}},\ \bibinfo {pages} {011013} (\bibinfo {year}
  {2017})}\BibitemShut {NoStop}%
\bibitem [{\citenamefont {Baxter}\ \emph {et~al.}(2011)\citenamefont {Baxter},
  \citenamefont {Dorogovtsev}, \citenamefont {Goltsev},\ and\ \citenamefont
  {Mendes}}]{baxter2011heterogeneous}%
  \BibitemOpen
  \bibfield  {author} {\bibinfo {author} {\bibfnamefont {G.~J.}\ \bibnamefont
  {Baxter}}, \bibinfo {author} {\bibfnamefont {S.~N.}\ \bibnamefont
  {Dorogovtsev}}, \bibinfo {author} {\bibfnamefont {A.~V.}\ \bibnamefont
  {Goltsev}}, \ and\ \bibinfo {author} {\bibfnamefont {J.~F.~F.}\ \bibnamefont
  {Mendes}},\ }\bibfield  {title} {\enquote {\bibinfo {title} {Heterogeneous
  k-core versus bootstrap percolation on complex networks},}\ }\href@noop {}
  {\bibfield  {journal} {\bibinfo  {journal} {Phys. Rev. E}\ }\textbf {\bibinfo
  {volume} {83}},\ \bibinfo {pages} {051134} (\bibinfo {year}
  {2011})}\BibitemShut {NoStop}%
\bibitem [{\citenamefont {Min}\ and\ \citenamefont
  {San~Miguel}(2018)}]{min2018competing}%
  \BibitemOpen
  \bibfield  {author} {\bibinfo {author} {\bibfnamefont {B.}~\bibnamefont
  {Min}}\ and\ \bibinfo {author} {\bibfnamefont {M.}~\bibnamefont
  {San~Miguel}},\ }\bibfield  {title} {\enquote {\bibinfo {title} {Competing
  contagion processes: Complex contagion triggered by simple contagion},}\
  }\href@noop {} {\bibfield  {journal} {\bibinfo  {journal} {Sci. Rep.}\
  }\textbf {\bibinfo {volume} {8}},\ \bibinfo {pages} {10422} (\bibinfo {year}
  {2018})}\BibitemShut {NoStop}%
\bibitem [{\citenamefont {Goltsev}\ \emph {et~al.}(2008)\citenamefont
  {Goltsev}, \citenamefont {Dorogovtsev},\ and\ \citenamefont
  {Mendes}}]{goltsev2008percolation}%
  \BibitemOpen
  \bibfield  {author} {\bibinfo {author} {\bibfnamefont {A.~V.}\ \bibnamefont
  {Goltsev}}, \bibinfo {author} {\bibfnamefont {S.~N.}\ \bibnamefont
  {Dorogovtsev}}, \ and\ \bibinfo {author} {\bibfnamefont {J.~F.~F.}\
  \bibnamefont {Mendes}},\ }\bibfield  {title} {\enquote {\bibinfo {title}
  {Percolation on correlated networks},}\ }\href@noop {} {\bibfield  {journal}
  {\bibinfo  {journal} {Phys. Rev. E}\ }\textbf {\bibinfo {volume} {78}},\
  \bibinfo {pages} {051105} (\bibinfo {year} {2008})}\BibitemShut {NoStop}%
\bibitem [{\citenamefont {Min}\ \emph {et~al.}(2014)\citenamefont {Min},
  \citenamefont {Do~Yi}, \citenamefont {Lee},\ and\ \citenamefont
  {Goh}}]{min2014network}%
  \BibitemOpen
  \bibfield  {author} {\bibinfo {author} {\bibfnamefont {B.}~\bibnamefont
  {Min}}, \bibinfo {author} {\bibfnamefont {S.}~\bibnamefont {Do~Yi}}, \bibinfo
  {author} {\bibfnamefont {K.-M.}\ \bibnamefont {Lee}}, \ and\ \bibinfo
  {author} {\bibfnamefont {K.-I.}\ \bibnamefont {Goh}},\ }\bibfield  {title}
  {\enquote {\bibinfo {title} {Network robustness of multiplex networks with
  interlayer degree correlations},}\ }\href@noop {} {\bibfield  {journal}
  {\bibinfo  {journal} {Phys. Rev. E}\ }\textbf {\bibinfo {volume} {89}},\
  \bibinfo {pages} {042811} (\bibinfo {year} {2014})}\BibitemShut {NoStop}%
\bibitem [{\citenamefont {Reis}\ \emph {et~al.}(2014)\citenamefont {Reis},
  \citenamefont {Hu}, \citenamefont {Babino}, \citenamefont {Andrade~Jr},
  \citenamefont {Canals}, \citenamefont {Sigman},\ and\ \citenamefont
  {Makse}}]{reis2014avoiding}%
  \BibitemOpen
  \bibfield  {author} {\bibinfo {author} {\bibfnamefont {S.~D.~S.}\
  \bibnamefont {Reis}}, \bibinfo {author} {\bibfnamefont {Y.}~\bibnamefont
  {Hu}}, \bibinfo {author} {\bibfnamefont {A.}~\bibnamefont {Babino}}, \bibinfo
  {author} {\bibfnamefont {J.~S.}\ \bibnamefont {Andrade~Jr}}, \bibinfo
  {author} {\bibfnamefont {S.}~\bibnamefont {Canals}}, \bibinfo {author}
  {\bibfnamefont {M.}~\bibnamefont {Sigman}}, \ and\ \bibinfo {author}
  {\bibfnamefont {H.~A.}\ \bibnamefont {Makse}},\ }\bibfield  {title} {\enquote
  {\bibinfo {title} {Avoiding catastrophic failure in correlated networks of
  networks},}\ }\href@noop {} {\bibfield  {journal} {\bibinfo  {journal} {Nat.
  Phys.}\ }\textbf {\bibinfo {volume} {10}},\ \bibinfo {pages} {762--767}
  (\bibinfo {year} {2014})}\BibitemShut {NoStop}%
\bibitem [{\citenamefont {Bianconi}(2018)}]{bianconi2018multilayer}%
  \BibitemOpen
  \bibfield  {author} {\bibinfo {author} {\bibfnamefont {G.}~\bibnamefont
  {Bianconi}},\ }\href@noop {} {\emph {\bibinfo {title} {Multilayer Networks:
  Structure and Function}}}\ (\bibinfo  {publisher} {Oxford University Press},\
  \bibinfo {year} {2018})\BibitemShut {NoStop}%
\bibitem [{\citenamefont {Chung}\ and\ \citenamefont
  {Lu}(2002{\natexlab{a}})}]{chung2002connected}%
  \BibitemOpen
  \bibfield  {author} {\bibinfo {author} {\bibfnamefont {F.}~\bibnamefont
  {Chung}}\ and\ \bibinfo {author} {\bibfnamefont {L.}~\bibnamefont {Lu}},\
  }\bibfield  {title} {\enquote {\bibinfo {title} {Connected components in
  random graphs with given expected degree sequences},}\ }\href@noop {}
  {\bibfield  {journal} {\bibinfo  {journal} {Ann. Comb.}\ }\textbf {\bibinfo
  {volume} {6}},\ \bibinfo {pages} {125--145} (\bibinfo {year}
  {2002}{\natexlab{a}})}\BibitemShut {NoStop}%
\bibitem [{\citenamefont {Chung}\ and\ \citenamefont
  {Lu}(2002{\natexlab{b}})}]{chung2002average}%
  \BibitemOpen
  \bibfield  {author} {\bibinfo {author} {\bibfnamefont {F.}~\bibnamefont
  {Chung}}\ and\ \bibinfo {author} {\bibfnamefont {L.}~\bibnamefont {Lu}},\
  }\bibfield  {title} {\enquote {\bibinfo {title} {The average distances in
  random graphs with given expected degrees},}\ }\href@noop {} {\bibfield
  {journal} {\bibinfo  {journal} {Proc. Natl. Acad. Sci. U.S.A.}\ }\textbf
  {\bibinfo {volume} {99}},\ \bibinfo {pages} {15879--15882} (\bibinfo {year}
  {2002}{\natexlab{b}})}\BibitemShut {NoStop}%
\end{thebibliography}
%

\end{document}